\documentclass[letterpaper,twocolumn,10pt]{article}
\usepackage{usenix-2020-09}

\usepackage{graphicx}
\usepackage{subcaption}
\usepackage{xspace}
\usepackage{xcolor}
\usepackage{boxedminipage,endnotes,balance}
\usepackage{pdfpages}
\usepackage{rotating}
\usepackage{colortbl}
\usepackage{transparent}
\usepackage[linesnumbered,ruled,vlined]{algorithm2e}
\usepackage{appendix}
\usepackage{multirow}
\usepackage{amsmath,amssymb,amsfonts}
\usepackage{relsize}
\usepackage{booktabs}
\usepackage{cite}
\usepackage{outlines}
\usepackage{cancel}
\usepackage{listings}
\usepackage{ifthen}
\usepackage{fancyhdr}
\usepackage{xurl}
\usepackage{bibspacing}
\usepackage{epsfig,comment}

\newcommand{\eg}{{e.g.,} }

\newcommand{\ie}{{i.e.}, }

\newcommand{\nop}[1]{}
\newcommand{\Fig}[1]{Fig.~\ref{fig:#1}\xspace}

\newcommand{\Tab}[1]{Table~\ref{tab:#1}\xspace}
\newcommand{\Sec}[1]{$\S$\ref{s:#1}\xspace}
\newcommand{\Subsec}[1]{$\S$\ref{ss:#1}\xspace}

\usepackage[font=small,labelfont=bf,textfont=bf,aboveskip=8pt]{caption}
\newcommand{\para}[1]{\smallskip\noindent\textbf{#1}~}

\newcommand{\cut}[1]{}

\newcommand{\name}{Mowgli\xspace}
\def\compactify{\itemsep=0pt \topsep=0pt \partopsep=0pt \parsep=0pt}
\let\latexusecounter=\usecounter

\newfont{\dft}{phvb at 6pt}
\newfont{\mft}{phvro at 6pt}
\newfont{\df}{phvb at 9pt}
\newfont{\mf}{phvro at 9pt}
  {\begin{list}{$\bullet$}%
     {\setlength{\parsep}{0pt}%
      \setlength{\topsep}{0pt}%
      \setlength{\itemsep}{0pt}}}%
  {\end{list}}

\makeatletter

\global\def\section{\@startsection {section}{1}{\z@}%
                                   {-1.5ex \@plus -0.8ex \@minus -.1ex}%
                                   {0.6ex \@plus.2ex}% changed this
                                   {\normalfont\bfseries\scshape\fontsize{11}{13}\selectfont}}
\global\def\subsection{\@startsection{subsection}{2}{\z@}%
                                     {-1.25ex\@plus -0.8ex \@minus -.1ex}%
                                     {0.3ex \@plus .1ex}% changed this
                                     {\normalfont\bfseries\fontsize{10}{12}\selectfont}}
\global\def\subsubsection{\@startsection{subsubsection}{3}{\z@}%
                                     {-1ex\@plus -1ex \@minus -.1ex}%
                                     {0.1ex \@plus .1ex}% changed this
                                     {\normalfont\itshape\fontsize{10}{12}\selectfont}}

\def\noeditingmarks{}
\newcommand{\textred}[1]{\textcolor{red}{#1}}
\ifx\noeditingmarks\undefined
   \newcommand{\pgwrapper}[2]{\textred{#1: #2}}
\else
   \newcommand{\pgwrapper}[2]{}
\fi

\newcommand{\squishlist}{
   \begin{list}{$\bullet$}
    { \setlength{\itemsep}{0pt}      \setlength{\parsep}{3pt}
      \setlength{\topsep}{3pt}       \setlength{\partopsep}{0pt}
      \setlength{\leftmargin}{1.0em} \setlength{\labelwidth}{1em}
      \setlength{\labelsep}{0.5em} } }
\newcommand{\squishend}{
    \end{list}  }

\SetKwInput{KwInput}{Input}                % Set the Input
\SetKwInput{KwOutput}{Output}              % set the Output

\begin{document}

\twocolumn[\begin{@twocolumnfalse}

\begin{centering}

{\Large \bf \name{}: Passively Learned Rate Control for Real-Time Video}

\def\refPrinceton{$^\P$}
\def\refuiuc{$^\dagger$}

{\vspace{0.3cm}
\large
\begin{tabular}[t]{cccc}
Neil Agarwal\refPrinceton{}&Rui Pan\refPrinceton{}&Francis Y. Yan\refuiuc{}
&Ravi Netravali\refPrinceton{}\\
\end{tabular}\par}

{\vspace{0.15cm}
\fontsize{11pt}{11pt}\selectfont
\begin{tabular}[t]{ccc}
\refPrinceton{}Princeton University&\refuiuc{}University of Illinois Urbana-Champaign\\
\end{tabular}}

\end{centering}

\vspace{1.2\baselineskip}

\end{@twocolumnfalse}]

\interfootnotelinepenalty 100000
\widowpenalty 100000
\clubpenalty 100000
\newfont{\tf}{phvro at 9.5pt}
\newfont{\tft}{phvro at 7.25pt}
\begin{sloppypar}
% \pagenumbering{gobble} % {gobble} to remove

\begin{abstract}
Rate control algorithms are at the heart of video conferencing platforms, determining target bitrates that match dynamic network characteristics for high quality. Recent data-driven strategies have shown promise for this challenging task, but the performance degradation they introduce during training has been a nonstarter for many production services, precluding adoption. This paper aims to bolster the \emph{practicality} of data-driven rate control by presenting an alternative avenue for experiential learning: leveraging purely existing telemetry logs produced by the incumbent algorithm in production. We observe that these logs contain effective decisions, although often at the wrong times or in the wrong order. To realize this approach despite the inherent uncertainty that log-based learning brings (i.e., lack of feedback for new decisions), our system, \name{}, combines a variety of robust learning techniques (\ie conservatively reasoning about alternate behavior to minimize risk and using a richer model formulation to account for environmental noise). Across diverse networks (emulated and real-world), \name{} outperforms the widely deployed GCC algorithm, increasing average video bitrates by 15–39\% while reducing freeze rates by 60–100\%.
\vspace{-3pt}
\end{abstract}
\section{Introduction}
\label{s:intro}

Real-time video conferencing is integral to our daily lives, with widespread use cases across many societal pathways including healthcare, education, gaming, and more. Key to their functionality are the rate control algorithms (e.g., Google Congestion Control or GCC~\cite{gcc}) that conferencing platforms employ. These algorithms are tasked with quickly (e.g., every 50 ms) characterizing network performance based on recent transmissions and selecting a target bitrate for the upcoming frames that maximizes content quality without introducing undue latency or stalls. This value is then shared with the local video codec which performs best-effort compression of the raw frames to match the target prior to transmission.

Recent years have witnessed a flurry of proposals for improving conferencing quality via improved rate control. Most notably, data-driven approaches have shown how reinforcement learning (RL)-based algorithms can substantially outperform GCC's hand-tuned variants~\cite{onrl,loki,concerto}. As in other networked-system domains~\cite{pensieve, puffer, orca, sage_cc}, the key is in making better use of dense application- and network-layer feedback signals to enable bitrate changes that more closely track rapid network fluctuations in the wild---a key challenge for GCC~\cite{gcc, concerto, onrl, loki}.

Yet, despite their promise, data-driven rate control algorithms have seen minimal adoption in practice. Our conversations with operators of large-scale video conferencing platforms reveal that the primary showstopper is the impact that these schemes have on the performance or quality of experience (QoE) observed by real users. Indeed, the experiential learning that these schemes pursue \emph{in real conferencing sessions} fundamentally involves exploring different rate decisions---both good and bad for QoE---and observing their effects in certain scenarios. Our experiments show that this trial-and-error process can increase video freeze rates up to $79\%$ and degrade video bitrates by up to $77\%$, yielding unacceptable QoE (\Subsec{motivations}). Data-driven strategies could forego such user-facing training by relying on simulators and emulators, but this may jeopardize efficacy in production settings due to the ``simulation to reality'' gap~\cite{sim2realitygap1, sim2realitygap2,puffer,onrl,r3net}.

This paper aims to build on recent data-driven rate control schemes, not by designing more performant algorithms, but instead by \emph{bolstering their practicality}. Our key insight is that the fine-grained telemetry logs that production conferencing platforms routinely capture for debugging and retrospective optimization~\cite{teams_dataset, zoom_logging, zoom_logging2, teams_logging, teams_logging2} already embed sufficient insight to guide data-driven algorithms to outperform their hand-tuned counterparts. The reason is that, while algorithms like GCC struggle to quickly match fluctuating network bandwidths, they 
often adjust target rates in the appropriate direction with delay (\Fig{gcc_shortcomings}). Consequently, we find that simply reorganizing the same decisions that GCC makes during video sessions (without exploring any alternate rate values) boosts conferencing bitrates by 19\% and decreases freezes by 80\% (\S\ref{ss:opportunity}).

We present \textbf{\name{}}, an end-to-end system that realizes such log-based learning to practically enhance rate control in video conferencing. \name{} starts by representing production telemetry logs that reflect the behavior and performance of a deployed algorithm (e.g., GCC) as more traditional (state, action, reward) tuples for RL. These logs then guide \name{}'s experiential learning, which follows the soft actor-critic algorithm~\cite{sac} to develop a lightweight neural network for rate control. Training occurs entirely offline using only logs (i.e., no video playback or simulation), and the resultant model is shipped to clients for deployment. 

Though conceptually straightforward, learning in this offline manner faces several challenges centered around \emph{uncertainty} (\S\ref{ss:design_challenges}). As noted above, the feasible learned improvements are rooted in discovering alternate sequences of logged actions that perform better in certain scenarios. However, assessing such sequences is fundamentally risky as they involve applying previously-seen actions to new scenarios without guarantees that the benefits will port, i.e., unlike traditional RL, we lack direct feedback for these new sequences. This limitation is worsened by the fact that logs in conferencing systems (1) typically reflect singular deterministic policies with little variation in action for a given scenario, and (2) are inherently noisy in that observed performance can be influenced not only by rate decisions, but also external phenomena such as codec behavior and stochastic network variations~\cite{concerto, salsify, puffer, sprout, verus}.

To manage uncertainty, \name{} incorporates two key techniques.
First, when estimating the consequence of an action in a given scenario, \name{} takes a \textit{conservative} approach---it lowers the estimated return if no similar state-action pair has been observed before, and proceeds only when the performance improvement outweighs the risk. Second, rather than estimating a single expected outcome, \name{} explicitly tackles environmental variance by learning a \textit{distribution} over all possible outcomes. This distributional perspective provides \name{} with richer insights, enabling more informed decision-making.
We detail the concrete algorithms in \Subsec{generator}.

We evaluated \name{} on a diverse set of emulated and real-world networks spanning 3G--5G cellular and wired broadband links. Overall, we observe that \name{} consistently outperforms GCC, increasing average video bitrates by 15--39\% while reducing freeze rates by 60--100\%. Further, \name{}'s wins closely mimic those of recent (impractical) online RL algorithms~\cite{loki, onrl, r3net}, falling within 0.5--13.1\% and 0--19\% for average bitrates and freeze rates, respectively.
\section{Background \& Motivation}
\label{s:background}

\subsection{Prior Work}
\label{ss:prior_work}
\para{Rule-based heuristics are suboptimal.}
Today's video conferencing applications (\eg Microsoft Teams, Google Hangouts,
Zoom) rely on rule-based algorithms to provide network bandwidth estimates
and guidance as to how to tune the encoding and sending bitrates over time.
A widely used, publicly available algorithm is Google Congestion Control (GCC)~\cite{gcc}. GCC first characterizes the current network usage based on recent packet delay and loss measurements, and then updates the target bitrate according to a fixed set of rules (\eg when the observed packet loss is less than 2\%, increase the
target bitrate by 5\%, or when the system is in a ``decrease'' state,
reduce the target bitrate by 15\%).

Google Congestion Control (and other rule-based heuristics) have been widely
observed to be suboptimal~\cite{concerto, onrl, loki, r3net, salsify}.
Recent reports have found that when using state-of-the-art rule-based heuristics,
over 20\% of over a million video conferencing sessions experience poor
performance~\cite{concerto}. Digging deeper, we find that GCC performs
particularly poorly in highly dynamic network conditions, where the available
bandwidth fluctuates (\eg in cellular networks). The general hardcoded set of
rules fails to fully leverage dense application/network signals that are necessary for predicting the appropriate bitrate adjustments in highly variable regimes. 

\begin{figure}[t]
    \centering
    \begin{subfigure}[b]{\columnwidth}
        \includegraphics[width=\textwidth]{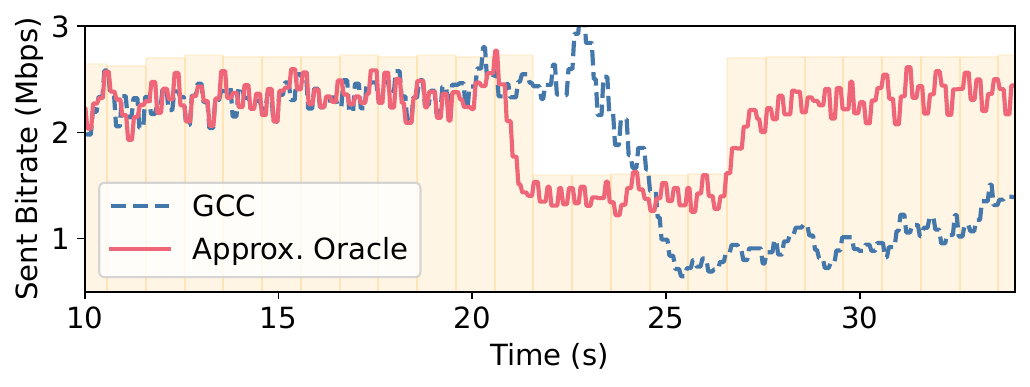}
        \vspace{-20pt}
        \caption{GCC overshoots network capacity after a bandwidth drop, causing video freezes.}
        \label{fig:gcc_short_a}
     \end{subfigure}
     \begin{subfigure}[b]{\columnwidth}
        \includegraphics[width=\columnwidth]{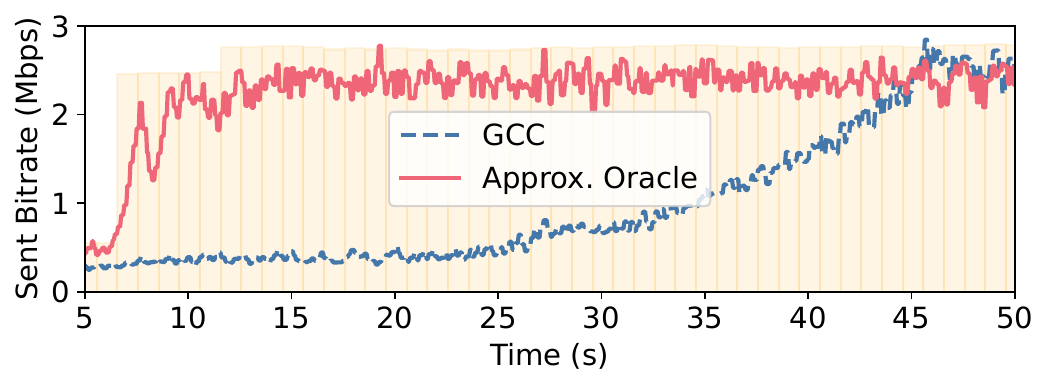}
        \vspace{-20pt}
        \caption{After an intermittent bandwidth drop, GCC ramps up slowly, leading to suboptimal bandwidth utilization.}
        \label{fig:gcc_short_b}
     \end{subfigure}
     \vspace{-15pt}
     \caption{Examples of GCC's pitfalls, which occur primarily in dynamic networks. To illustrate potential performance improvement opportunities, we plot an oracle algorithm.}
    \label{fig:gcc_shortcomings}
\end{figure}

As a result, for example, when bandwidth drops, GCC can fail to rapidly and
appropriately adjust the bandwidth, unnecessarily overshooting the
available network capacity and incurring video freezes
(\Fig{gcc_short_a}). Further, after an intermittent drop in
bandwidth, GCC can be unnecessarily delayed in fully ramping up to the available
bandwidth (\Fig{gcc_short_b}). As quality expectations for video
conferencing steadily increase with its growing ubiquity, there is a pressing
need to find improved rate control algorithms over today's
rule-based heuristics.

\para{Data-driven approaches are promising.}
In response, the community has explored data-driven approaches to
generate alternative rate control algorithms (\eg R3Net~\cite{r3net}, OnRL~\cite{onrl}, Loki~\cite{loki}). 
Such machine learning-based approaches are particularly skilled at extracting
patterns from highly-dense information and have been shown to deliver nontrivial
improvements over today's state-of-the-art rule-based heuristics. Given the
challenge of explicitly identifying the exact action an algorithm should take in
each scenario and using standard supervised learning techniques to learn that
mapping, recent proposals have opted to leverage reinforcement learning (RL).
An RL agent learns by interacting with an environment and
leveraging feedback to iteratively adjust its behavior to
maximize cumulative reward over a horizon. In rate control for
video conferencing, the RL agent outputs bitrate updates and leverages the
transport/application layer feedback to update its decision-making policy.

\subsection{Motivations}
\label{ss:motivations}
As confirmed by prior work~\cite{onrl, loki, r3net} and our results in
\Sec{eval}, data-driven methods can deliver significant wins over rule-based
heuristics and are crucial to enabling the effective rate control that
today's video conferencing applications demand. Surprisingly, despite their
potential benefits, these solutions have yet to gain traction in production
deployments. In conversations with the operators of major production video
conferencing deployments, these solutions do not meet the practicality
constraints of production environments. In particular, the biggest concern is
the disruption of client video conferencing sessions during model training.

\begin{figure}[t]
    \centering
    \includegraphics[width=\columnwidth]{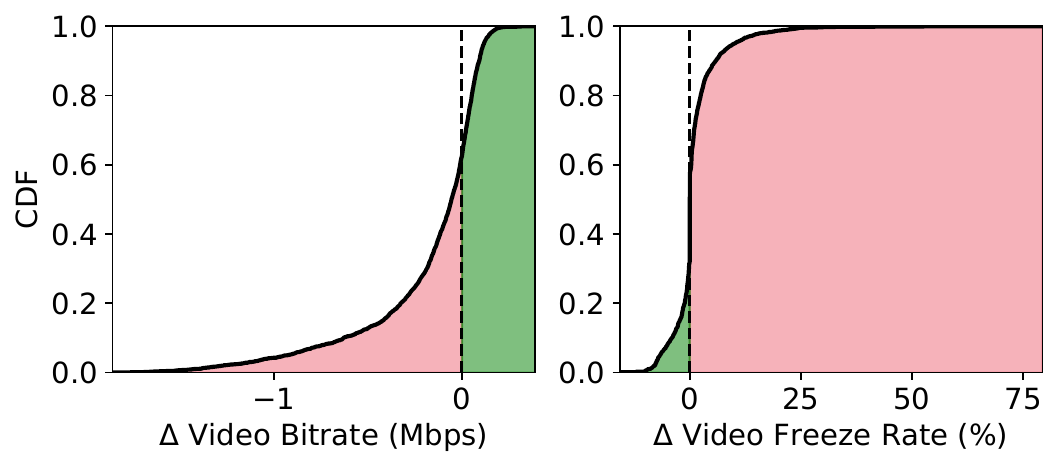}
    \vspace{-20pt}
    \caption{Distribution of changes to video QoE observed (relative to GCC) during online RL model training. Improvements are shaded in green; degradations are shaded in red.}
    \label{fig:online_qoe_degradations}
\end{figure}

Fundamental to existing solutions is their use of reinforcement learning, a
trial-and-error learning mechanism that trains a model by iteratively interacting
with an environment and updating the model based on the interaction outcomes.
For existing solutions, randomly initialized RL agents (models) are
placed directly on client devices and are allowed to dictate the bitrate
decisions for real user-facing video conferencing sessions. Over time, the
models gain experience and converge to a high-performing policy; however, this
comes at the cost of disruption to the quality of video conferencing calls
during the training process. Prior approaches explicitly encourage exploration
of different actions and behavior during training (by adding an entropy
bonus to the learning objective); the bitrate decisions taken during exploration
may be far from the ideal bitrate decision for that particular scenario and
dramatically degrade the quality of a video call.

\para{Quality disruptions of online learning.} To observe how the performance of video conferencing sessions is affected
during training, we ran a series of experiments. We train an online
RL algorithm in our video conferencing testbed and periodically
record the QoE metrics of affected sessions.
More details about our video conferencing testbed, online RL implementation, and QoE
metrics can be found in \Sec{eval}. Note that our implementation includes the
temporary fallback mechanism introduced by OnRL~\cite{onrl} which switches back
to the rule-based heuristic when overuse is detected in the hope of reducing
catastrophic behavior.

\begin{figure}[t]
    \centering
    \includegraphics[width=\columnwidth]{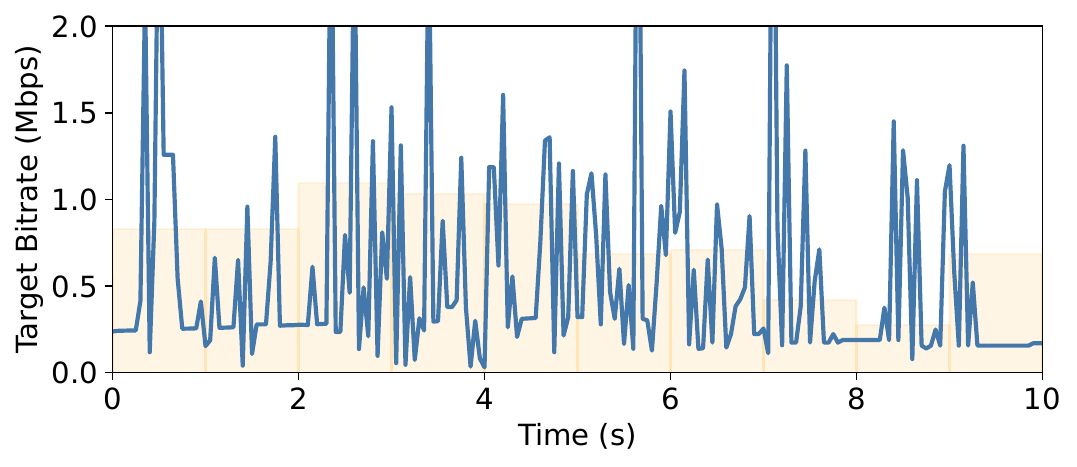}
    \vspace{-20pt}
    \caption{Example disruptive behavior observed during online RL training;  network bandwidth is shaded in yellow.}
    \label{fig:online_exploration_examples}
\end{figure}

\Fig{online_qoe_degradations} shows the effects on performance (relative to GCC) during model training for two key QoE metrics. We find that 62\% of calls experience a worse average video bitrate, with degradations as low as \textminus 1.9 Mbps (for context, the average bitrate of a call is 1.03 Mbps). 43\% of calls experience higher video freeze rates, with freeze rate increases as much as +79\%. Intuitively, these performance degradations are due to the exploratory nature of online RL mechanisms. \Fig{online_exploration_examples} highlights some of the disruptive behavior that causes performance degradations: repeated switching between low and high bitrates, bandwidth underutilization, and overaggressive bitrate ramp-ups. Fallback mechanisms fail to resolve these issues because they are only activated once catastrophic behavior is detected. Further, these degradations are exacerbated by the iterative development process of model optimization across different model designs and hyperparameters---the costs of training (\eg QoE degradations) will be incurred every time.

\subsection{Alternative Approaches}
\label{ss:alternative_approaches}

\para{Training in simulation.} Earlier work~\cite{onrl, r3net, concerto} has investigated training policies in simulated or emulated environments to avoid the user-facing quality degradations seen in previous approaches. However, they found that these models perform poorly in production due to the ``simulation-to-reality'' gap~\cite{sim2realitygap1, sim2realitygap2}, where simulated environments fail to fully and accurately capture the complexity of production systems and any slight deviation in system dynamics can lead to cascading performance issues. For more details, we refer readers to prior reports~\cite{onrl, r3net}. Improving simulation fidelity in networked systems remains an active area of research~\cite{causalsim, veritas}.

This gap also explains why starting with a ``pre-trained model'' from simulation and then fine-tuning it online in production is problematic. The disparity between simulated and real environments, combined with the significant experience required for fitting an RL model to a new distribution of environments, means the model typically needs major revisions rather than minor adjustments. This would disrupt production services and negate the benefits of ``pre-training''. Further, in our review of related work~\cite{onrl, r3net, loki}, we found that prior approaches always trained models from scratch in the deployment setting, starting with completely untrained models.

\para{Tuning GCC knobs.} Prior work~\cite{oboe} has shown the benefits of using data-driven parameter tuning for ABR heuristic algorithms. However, applying this approach to GCC faces three main challenges. First, GCC bakes in handcrafted rules that are inherently heuristic and suboptimal, thereby constraining the potential impact of hyperparameter tuning. Second, GCC does not support extracting insights from a richer set of network signals, depending instead primarily on a single, unreliable network signal (the gradient of queuing delay). Lastly, its codebase includes hundreds of tunable knobs, making it unclear how to prioritize or manage them effectively.

\subsection{Summary} Recent data-driven approaches demonstrate considerable improvements over their rule-based counterparts. However, they face minimal adoption in production due to the disruptive nature of their interactive trial-and-error-based online learning approach. Alternatively, policies trained in simulation perform poorly when deployed to production due to the ``simulation to reality'' gap. How to capitalize on the performance improvements offered by data-driven approaches while adhering to the constraints of production environments is an open question faced by video conferencing system operators.

\vspace{-1pt}

\section{Vision: Learning By Observation}
\label{s:vision}
In this paper, we claim that there is a viable path forward, but it requires rethinking the end-to-end design of data-driven rate control systems, from how we source the data to how we leverage and learn from it.

\vspace{-1pt}

\subsection{The Data of Prior Approaches}
\label{ss:prior_data}
Prior data-driven approaches collect training data by iteratively deploying and continually updating DNN-based rate control policies in production video conferencing systems, and logging the resulting behavior and outcomes~\cite{onrl, loki, r3net}. These logs are shipped from users to a central server, where they are processed and then fed to a training algorithm that updates the weights underlying the current version of the rate control policy. The updated weights are shipped out to the clients and the next round of training proceeds. 

At the central server, data processing of the logs involves extracting structured sequential data, \ie series of (state, action, reward) tuples. A ($s_t$, $a_t$, $r_t$) tuple represents the following: at a given time $t$, the rate control policy takes as input recently observed information about application and transport layers (captured as vector $s_t$) and outputs an updated target bitrate $a_t$. The effect of updating the target bitrate to $a_t$ on application performance (\eg the change in video throughput, freezes, frame delay, etc.) is quantified as $r_t$. Such sequences of (state, action, reward) tuples offer a structured way to reflect the experiences and effects of a rate control policy.

This data provides critical feedback to the training process by reinforcing decision-making that leads to good behavior (high reward) and penalizing decision-making that leads to poor behavior (low reward). Recall that the training algorithm initially knows nothing about the environment or the ideal behavior. As a result, the initial rate control policies take random actions. This exploratory behavior helps build a growing set of diverse experiences, each trajectory corresponding to a different sequence of possible decisions. Over time, as the weights of the rate control policies are updated and additional logs are collected, the model identifies high-potential paths and ultimately, converges to a policy that maximizes the cumulative reward. In the process, each step of deployment and data collection provides a critical feedback loop, enabling the training algorithm to (1) test out new behaviors, and (2) correct any misunderstandings in the learned model.

However, as quantified in \Subsec{motivations}, accumulating this rich and diverse dataset requires deploying partially trained policies to production environments and disrupting the quality of video conferencing sessions for users. From the perspective of a production deployment operator, this is not a viable option.

\subsection{An Alternative Source of Data} 
\label{ss:alternate_source}

In this paper, we posit that there is a viable alternative source of data: the experiences of the rule-based algorithm \textit{currently deployed} in production settings for rate control (\eg Google Congestion Control). Unlike prior approaches, this data source can be obtained without deploying partially trained or untested rate control policies, thereby avoiding the disruptions outlined in \Subsec{motivations}. Instead, the data is collected from a rate control heuristic already in production, where it is logged at a fine granularity for purposes such as monitoring, debugging, and improving the heuristic algorithm~\cite{teams_dataset, zoom_production_logging}. Similar to previous methods, these logs can be post-processed to extract sequences of (state, action, reward) tuples, capturing the experiences of the deployed rate control policy.

However, ultimately, these logs reflect the behavior of a single, static rate control policy (production systems typically deploy a single algorithm, \eg the in-house rule-based heuristic). The logs of prior approaches, on the other hand, reflect the behavior of hundreds of rate control policies, learned and deployed over hundreds of epochs of training. As previously described, the diversity in behaviors enables the algorithm to learn an effective rate control policy. This begs the question: how we can learn a better rate control policy from the experiences of a single rate control policy?

Besides the interactive RL approaches of prior work, there exist alternative data-driven approaches such as imitation learning (\eg behavior cloning)~\cite{imitation_learning}. However, the goal of these methods is to learn the \emph{same} behavior as reflected in the training data; our goal is to learn a \emph{better} rate control strategy.

\subsection{Opportunity}
\label{ss:opportunity}

\begin{figure}[t]
    \begin{subfigure}[b]{\columnwidth}
        \includegraphics[width=\textwidth]{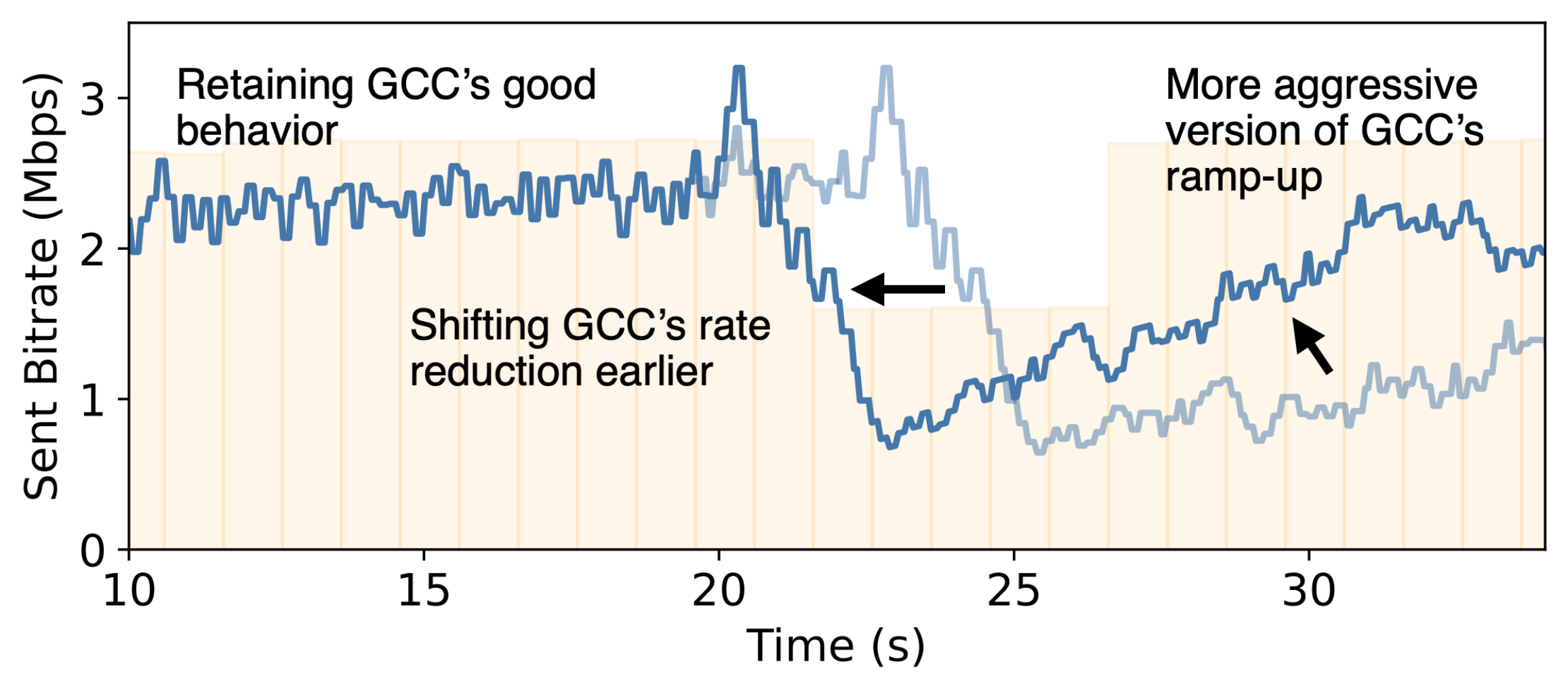}
        \vspace{-20pt}
        \caption{After the network bandwidth drops at \texttt{t=22}, it takes $3$ seconds for GCC's sending bitrate to catch up. By shifting GCC's bitrate reduction earlier, we can reduce the degree of network overutilization.} 
        \label{fig:stitching_intuition_1}
     \end{subfigure}
     \begin{subfigure}[b]{\columnwidth}
        \includegraphics[width=\columnwidth]
        {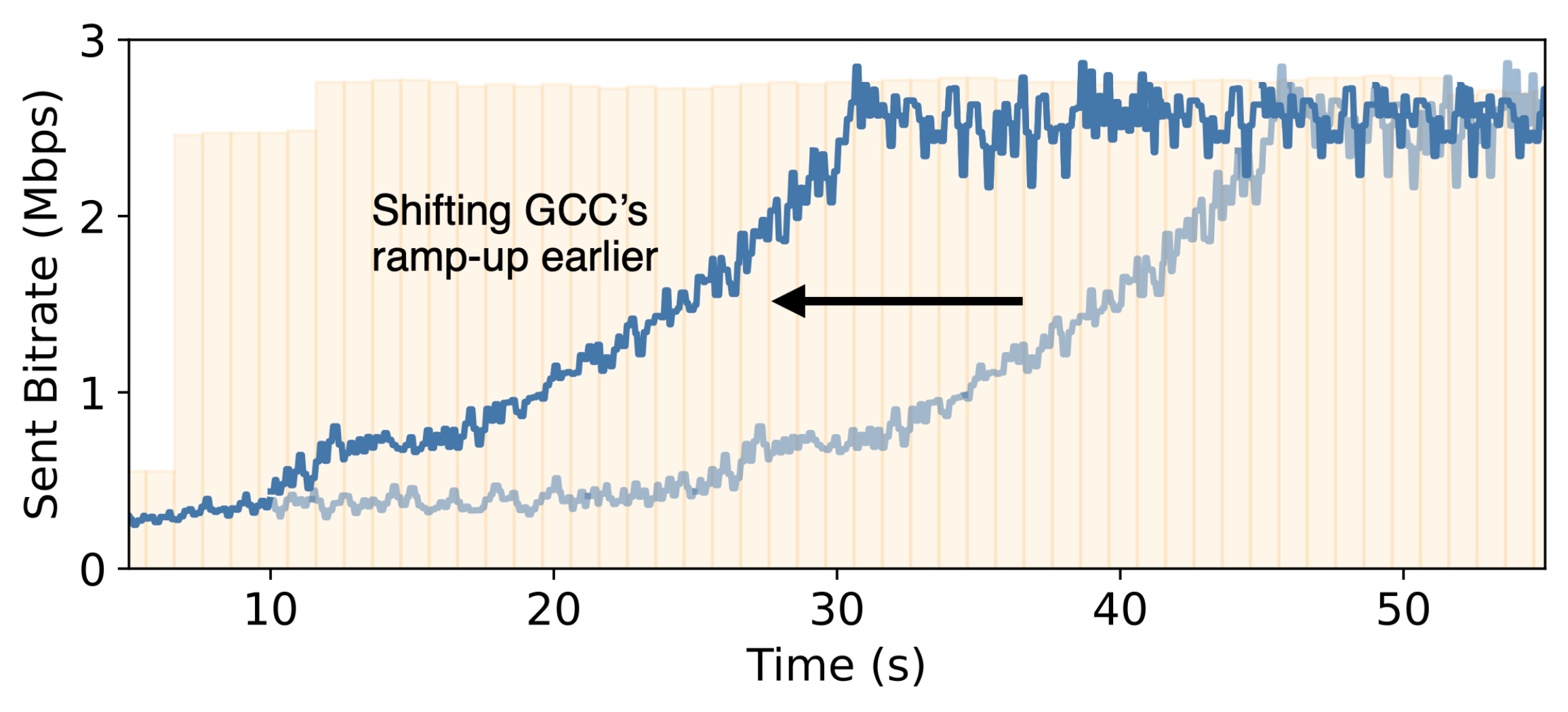}
        \vspace{-20pt}
         \caption{After the network bandwidth increases at \texttt{t=7}, it takes $40$ seconds for GCC's sending bitrate to catch up. By shifting GCC's bitrate ramp-up earlier, we can increase network utilization.}
        \label{fig:stitching_intuition_2}
     \end{subfigure}
     \caption{Potential improvements by rearranging sequences of actions within GCC's logs.}
    \label{fig:stitching_intuition}
\end{figure}

In this section, we introduce a novel approach to learning from the logs of GCC; we start by describing a key observation about the nature of GCC's shortcomings.

As discussed in \Subsec{prior_work}, GCC’s suboptimal performance stems from delayed or overly cautious responses to bandwidth changes. During sudden bandwidth drops (\Fig{stitching_intuition_1}), GCC's reliance on hand-tuned thresholds of limited network signals delays congestion detection and subsequent corrective actions. This lag results in network overutilization and video freezes. When network capacity increases, GCC exhibits overly cautious ramp-up behavior, hesitating to increase the bitrate for fear of overshooting the available capacity (\Fig{stitching_intuition_2}).

Notably, despite these issues, \textit{GCC eventually makes the correct directional adjustments}--- ramping down after detecting congestion or gradually increasing its bitrate when the network allows. This observation is crucial: it implies that, within the very logs where GCC performs poorly, there are also \textit{positive} examples that demonstrate how timely or more aggressive actions could have prevented unnecessary freezes or underutilization. These examples present a unique opportunity to develop an improved rate control algorithm directly from GCC’s logs. Instead of considering entirely new actions or behaviors, we can improve through careful selection and application of actions already demonstrated in the logs.

For example, in \Fig{stitching_intuition_1}, as GCC eventually reduces its bitrate following a bandwidth drop, the trace captures a range of actions: initially, actions at the same bitrate and subsequently, actions at lower bitrates. These logs also capture detailed network telemetry data reflecting the state before and after each action. Together, this data provides valuable examples that can be analyzed and optimized. Specifically, these examples can be used to (1) identify congestion signals more effectively and (2) select lower target bitrates more promptly. Similarly, as illustrated in \Fig{stitching_intuition_2}, when GCC (conservatively) increases its bitrate in response to improved bandwidth, the telemetry traces capture a sequence of steadily increasing bitrate adjustments. These actions, paired with their corresponding telemetry signals, provide examples of different responses and ultimately, a path to achieving improved network utilization.

Both scenarios illustrate that, despite capturing suboptimal decisions, GCC’s logs contain valuable data for achieving critical improvements. By systematically analyzing GCC’s behavior alongside dense telemetry signals, it becomes possible to reorder its actions into a more optimal sequence without the need to devise entirely new strategies. The central insight lies in leveraging GCC’s imperfect but fundamentally directionally correct decision sequences as foundational building blocks for constructing a more effective rate control policy.

To quantify the potential gains from this log-based learning approach, we implemented an approximate oracle algorithm that has access to ground-truth network dynamics but is restricted to the set of actions that appear in a given GCC log. On the network trace depicted in \Fig{stitching_intuition_1}, the oracle achieves a 52\% increase in video bitrate and a 98\% reduction in freeze rates. For the scenario in \Fig{stitching_intuition_2}, it achieves an 80\% increase in video bitrate and a 79\% reduction in freeze rates. Across our entire corpus of network traces (\Sec{eval}), this oracle-based approach yields a 19\% improvement in video bitrate and an 80\% decrease in freeze rates compared to GCC.

\vspace{-4pt}
\subsection{Challenges}
\label{ss:design_challenges}

However, realizing this approach in a way that is both practical and effective while adhering to the constraints of a production deployment setting requires addressing two key challenges.

\para{Challenge \#1: Lack of feedback.}
Ultimately, the goal is to learn an improved rate control policy without disrupting users (the main pitfall of prior approaches). However, producing an improved rate control policy requires learning a strategy that deviates from the behavior seen in the GCC logs (\eg an alternative sequence of actions). Learning this alternative (and better) strategy requires reasoning about (and extrapolating) the expected outcomes of alternative behaviors. This is a risky proposition without access to feedback (\ie testing the new strategy and validating the prediction). The greater the deviation, the greater the potential risk of extrapolation error; any errors in extrapolating will compound, as the resulting deviation will lead to more deviation. Ultimately, this comes at the detriment of the performance of the learned policy. In the field of learning sequential decision-making without feedback, this phenomenon is referred to as ``distribution shift''~\cite{levine_offline_rl_tutorial}. In our evaluation ablation studies (\Subsec{ablation}), we find that failing to address distribution shift can dramatically cause performance degradations, increasing P90 video freeze rates over $12\times$. To address this issue in the context of learning improved rate control policies, we need a way to effectively balance decision-making deviations with risk mitigation.

\para{Challenge \#2: Environmental variance.}
Further complicating the ability to develop an understanding of how any given target bitrate update affects observed outcomes (rewards) is the presence of external phenomena outside the control of the bitrate decision-making policy that affect application behavior and ultimately, the outcomes of a given bitrate update. Unlike prior work~\cite{concerto, r3net} that leverages emulation or simulation-based systems and can control for (or eliminate) the amount of noise introduced, production-based deployments do not have that luxury. Concretely, we find that this noise manifests in two ways. First, video conferencing applications apply additional downstream application logic after consuming a target bitrate update from the bitrate controller; this logic affects the \emph{achieved} encoding and sending bitrate~\cite{concerto, salsify}. Second, network conditions change rapidly (often in unpredictable ways~\cite{puffer, sprout, verus})---taking the same action at the same state in two different instances could result in different outcomes because of rapid changes in network conditions (\ie the bandwidth dropped). Ultimately, this makes it difficult to determine whether differences in the observed outcomes of two bitrate decisions arise from external factors or the effectiveness of the target bitrate decision itself. In our evaluation studies (\Subsec{ablation}), we find that failing to address the effects of environmental variance can increase P90 video freeze rates over 10$\times$. To have an effective solution, we need a way to explicitly express and account for environmental noise.
\section{Design}
\label{s:design}

\begin{figure}[t]
    \centering
    \includegraphics[width=\columnwidth]{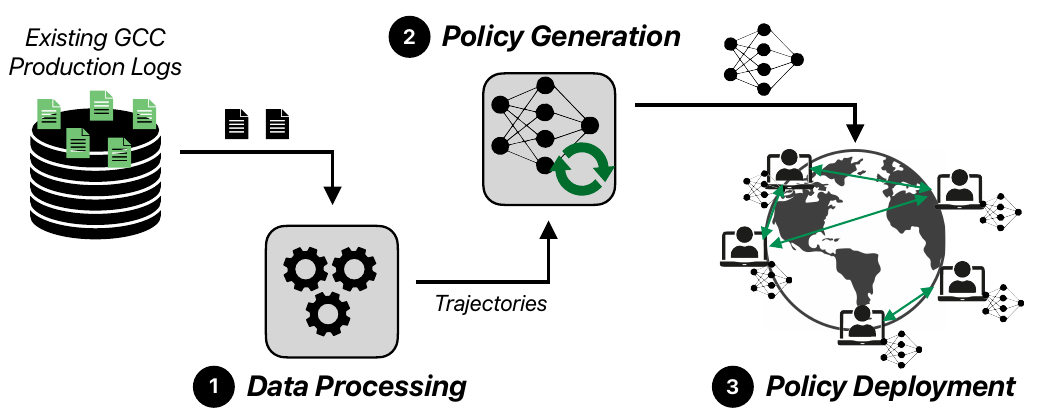}
    \caption{Overview of \name{}'s 3 phases: (1) data processing for extracting trajectories from GCC telemetry logs, (2) policy generation for model training, and (3) policy deployment.}
    \label{fig:system}
\end{figure}

\Fig{system} describes \name{}'s workflow. First, \name{} consumes existing production logs or builds on existing logging instrumentation to extract telemetry signals and generate trajectories of (state, action, reward) tuples that reflect the experiences of the underlying rate control algorithm (\Subsec{dataprocessing}). Then, \name{} uses these trajectories to train a lightweight neural network for rate control (\Subsec{generator}). Training occurs entirely offline using only telemetry logs, \ie no video playback or simulation is required; the resulting policy is then deployed on client devices (\Subsec{deployment}). We describe implementation details in \Subsec{implementation}.

\subsection{Data Collection \& Processing}
\label{ss:dataprocessing}

\name{} operates on production telemetry data that reflects the experiences of in-house rate control algorithms. These logs are typically already captured for other purposes such as debugging, observability, and quality assurance~\cite{zoom_production_logging, teams_dataset}. For example, recently released logs from Microsoft Teams capture periodic (every 60 ms) application and transport layer metrics (\eg packet loss, packet delay, received bitrate, and target bitrate) along with session-level QoE metrics (\eg duration of video freezes, average media bitrate received)~\cite{teams_dataset}. If the existing production logs do not provide the necessary data, we can build on existing instrumentation platforms to extract the additional data (\eg Zoom, Microsoft Teams, WebRTC have built-in logging capabilities~\cite{zoom_logging, zoom_logging2, teams_logging, teams_logging2, webrtc_logging, teams_dataset}). Once logs are aggregated across clients to a central server, we apply \name{}'s processing logic to extract trajectories of (state, action, reward) tuples.

We define the state vector and action based on prior work~\cite{onrl, loki, r3net}. The state vector is a window of periodically captured transport and application-level statistics (we use a window of 1 second). We describe the state vector in \Tab{features}. We find that augmenting the state vector inputs with four additional features further improves performance: the previous bitrate action, the minimum RTT observed so far, the number of timesteps since the last transport feedback report was received, and the number of timesteps since the last packet loss report was received. We measure the impact of these additional features in \Subsec{ablation}. The action is the updated target bitrate that is consumed by the application. We define the reward as a function of the achieved bitrate, the average RTT, and the average packet loss. Inspired by prior work, we leverage the following formulation:
\vspace{-2pt}
\begin{equation}
    R = \alpha \cdot throughput - \beta \cdot delay - \gamma \cdot loss
\vspace{-2pt}
\end{equation}

We normalize throughput to the range (0, 6 Mbps) and the delay to (0, 1000 ms); we set $\alpha$ to 2, $\beta$ to 1, and $\gamma$ to 1.

\begin{table}[t]
    \setlength\arrayrulewidth{1pt}
    \centering
\begin{tabular}{l}
\toprule
\multicolumn{1}{c}{\textbf{State Vector Inputs}}        \\
\midrule
Sent Bitrate                                   \\ 
Acknowledged Bitrate                           \\ 
Previous Action                                \\ 
One-Way Packet Delay                           \\ 
One-Way Packet Delay Jitter                    \\ 
Inter-packet Arrival Delay Variation           \\ 
Round Trip Time                                \\ 
Minimum Round Trip Time Observed So Far                   \\ 
Timesteps since Last Transport Feedback Report \\ 
Packet Loss                                    \\ 
Timesteps since Last Packet Loss Report \\
\bottomrule
\end{tabular}
\vspace{-5pt}
\caption{State vector of transport and application-layer statistics logged every $\sim 50$ ms.}
\label{tab:features}
\end{table}

\subsection{Policy Generation}
\label{ss:generator}
Next, \name{} leverages the aggregated and preprocessed logs to generate improved rate control policies.
In \Sec{vision}, we described an approach to improve upon GCC by rearranging the actions within a GCC log. In this section, we explain how to realize this approach in practice. First, we describe \name{}'s use of a lightweight neural network to represent a rate control policy on client devices; we then detail how we overcome the challenges of realizing this approach presented in \Subsec{design_challenges}.

\para{Leveraging neural networks.} 
To reason about potential actions for a scenario, we need a way to estimate the expected outcomes for (state, action) pairs. We turn to neural networks that we can train offline (using GCC logs) and subsequently deploy to client devices. In particular, we leverage a state-of-the-art learning algorithm, Soft Actor-Critic (SAC)~\cite{sac}.

SAC consumes the trajectories of (state, action, reward) tuples previously experienced by GCC and trains two complementary functions (both represented by parameterizable neural networks): the actor and the critic. The actor network ($\pi_{\theta}$) learns a deterministic policy that maps states to actions; the critic network ($Q_\phi$) evaluates the expected long-term reward (return) for a given (state, action) pair. SAC leverages the critic to provide a learning signal (via the Q-value) for the actor, enabling it to improve the policy by following the gradient of expected returns.  

Following standard Q-Learning~\cite{watkins1992q} techniques, we train the critic by minimizing the Mean Squared Bellman Error:
\vspace{-5pt}
\begin{equation}
\mathbb{E}_{(s_t,a_t,r_t)\sim{D}}\Bigl[(Q_\phi(s_t,a_t)-(r_t + \gamma \cdot \max_{a’} Q_\phi(s_{t+1},a’)))^2\Bigr]
\label{eq:critic_loss}
\vspace{-5pt}
\end{equation}

The actor network is updated by maximizing the Q-value predicted by the critic: 
\vspace{-5pt}
\begin{equation}
\max_\theta \mathbb{E}_{s\sim{D}} [Q_\phi(s, \pi_\theta(s))]
\label{eq:solve_pi}
\vspace{-5pt}
\end{equation}

We train $Q_\phi$ and $\pi_\theta$ in tandem, repeatedly iterating through the dataset of (state, action, reward) tuples. The pseudocode in Algorithm~\ref{alg:complex} describes the key components of the dual training process: updating the critic, and then updating the actor. Upon convergence (and during inference), we only need to retain $\pi_\theta$: given state $s$, return $\pi_\theta(s)$.
\vspace{-5pt}

\begin{algorithm}
\DontPrintSemicolon
\KwInput{Corpus of GCC Observations: $D \in \langle S, A, R \rangle$}
\SetKwProg{Fn}{def}{:}{}
\SetKwProg{Repeat}{repeat until convergence}{:}{}
\SetKwFunction{FUpdate}{update\_bitrate}

\Repeat{}{
    Randomly sample a batch of transitions, $B=\{(s,a,r,s')\}$ from $D$ \;
    Compute target: \;
    \qquad $y \gets r + \gamma \cdot Q_\phi(s', \pi_\theta(s'))$ \;

    Update critic function: \;
    
    \qquad $\nabla_\phi \frac{1}{|B|} \sum_{(s,a,r,s')\in{B}} (Q_\phi(s,a)-y)^2$ \;

    Update actor network: \;
    
    \qquad $\nabla_\theta \frac{1}{|B|} \sum_{s\in{B}} Q_\phi(s, \pi_\theta(s))$ \;
} 

\Fn{\FUpdate{state s}}{
    $a \gets \pi_\theta(s)$ \;
    \KwRet $a$;
}
\caption{Soft Actor Critic Algorithm}
\label{alg:complex}
\end{algorithm}

Further, we prepend the actor and critic networks with a learned embedding over the raw state vector to take advantage of the temporal aspect of the state. Following prior work, the state is a windowed series of metrics---we incorporate a Gated Recurrent Unit (GRU) to extract trends over the window and reduce the raw state to a more condensed vector~\cite{gru}. Implementation details can be found in \Subsec{implementation}.

\para{Conservative learning.} While the rate control policies are now trained offline in the cloud, the challenge of a lack of feedback persists \textbf{(Challenge \#1)}. Learning an improved rate control policy requires learning a strategy that deviates from the behavior seen in GCC. The greater the deviation, the greater the potential risk of extrapolation error; any errors in extrapolating will compound and drastically harm performance. 
In this section, we describe how the errors come to be in the context of the actor and critic networks, and then we go on to describe a risk mitigation strategy.

Recall that the learned critic function ultimately guides the decision-making policy (actor) to select actions that optimize for the best behavior. The critic $Q_\phi$ is responsible for learning a regression to estimate the value of (state, action) pairs observed in the dataset. It is also responsible for learning a regression to estimate the value of (state, action) pairs \textit{not seen} in the dataset. The latter is much harder because it requires extrapolating---that is, using values of observed (state, action) pairs to estimate values for unseen (state, action) pairs. As a result, the regressor is likely to be more error-prone for those unseen regions of the state-action space. This is particularly problematic because the actor is trained to learn actions that maximize the critic function (Equation~\ref{eq:solve_pi}). Any mistakes or erroneous value assignments can lead the actor astray; in particular, the actor becomes biased toward selecting actions that have been erroneously assigned high values. The result is a poor-performing policy, one that can be potentially worse than the one featured in the data used to train it.

To address this, we opt for the following approach: when leveraging the output of the learned critic to teach the actor, trust the estimates more when the regressor is confident (\ie the estimate is based on (state, action) pairs seen or are close to those observed in the dataset) and trust the estimates less when the regressor is less confident (\ie the estimate is heavily extrapolated). Recall that in cases where GCC performs poorly, it eventually makes adjustments in the correct direction (\Subsec{opportunity})---these sequences provide a sufficient number of examples to confidently extrapolate about alternate trajectories in areas that GCC needs improvement. 

In practice, we can achieve this by penalizing the estimated values for low-confidence regions and elevating those for high-confidence regions. Consequently, when the actor is leveraging the output of the critic, it learns to select actions that maximize the modified estimated value instead (\ie one that takes into account the accuracy of the estimate) and ultimately, avoid falling into a trap of taking actions with erroneously high estimates due to errors in the learned regressor.

In particular, we leverage a state-of-the-art technique known as Conservative Q-Learning (CQL)~\cite{cql}. CQL adds a regularizer to the critic's loss function (Equation \ref{eq:critic_loss}):
\vspace{-2pt}
\begin{equation}
\alpha \cdot \mathbb{E}_{s\sim{D}} \Bigl[\mathbb{E}_{a\sim{\pi(a|s)}}Q(s,a)-\mathbb{E}_{a\sim{D}}Q(s,a)\Bigr] \;
\label{eq:cql_regularizer}
\vspace{-2pt}
\end{equation} 

The regularizer first guides the critic to learn a lower bound on estimated values for all (state, action) pairs (\ie a ``conservative'' estimate); simultaneously, it ``pushes up'' the values for (state, action) pairs observed in the training dataset. CQL provides an adjustable parameter $\alpha$ to identify the appropriate magnitude of the conservative penalty: too high of a penalty creates a conservative, low-risk policy that performs similarly to the behavior featured in the logs; too low of a penalty negates the benefits of this approach altogether, resulting in a risky (and potentially, low-performing) policy. We empirically find that setting $\alpha=0.01$ achieves the best tradeoff; in our ablation studies in \Subsec{ablation}, we compare different settings of $\alpha$.

\para{Distributional representation.}
The second main challenge of learning from heuristic data collected in production deployment settings is the presence of external phenomena outside the control of the rate control policy that ultimately complicates reasoning about the observed outcomes for a given (state, action) pair \textbf{(Challenge \#2)}. To address this, we opt to explicitly account for the variance in expected return for a given (state, action) pair by modifying the way the learned value is represented. Traditionally, the critic learns a scalar value---\ie the expected value of starting at state $s$ and taking action $a$. We modify the critic function to learn a \emph{distribution} over expected outcomes instead~\cite{distributional_rl}. A probability distribution explicitly accounts for the possibility of multiple different futures, despite the bitrate agent taking the same action. Directly encoding a probability distribution into the model results in a more detailed representation of the environment's stochasticity and a more comprehensive way of capturing the effects of any given decision. 

To implement this, we modify the output of the critic function to be a vector (representing a distribution) and update the loss function to support a vector representation. In line with prior work, we incorporate the Quantile Huber loss function, which compares distributions and penalizes estimates differently depending on which quantile they belong~\cite{quantile_regression}. 

\subsection{Policy Deployment}
\label{ss:deployment}

\begin{figure}[t]
    \centering
    \includegraphics[width=1.0\columnwidth]{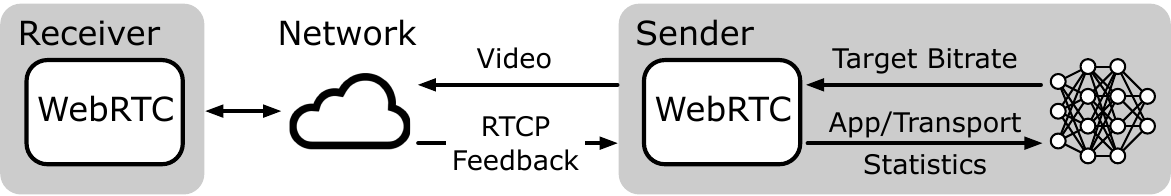}
    \vspace{-15pt}
    \caption{\name{} deploys its model in the sender application's rate control logic.}
    \label{fig:policy-deployment}
\end{figure}

Based on insights from our generalization study (\Subsec{generalization}), \name{} achieves performance wins across diverse network scenarios, provided that the corresponding state/action distributions are represented in the consumed telemetry logs. To adapt to new network environments, \name{} continuously monitors these logs, and if a shift in the underlying state/action distribution is detected, the system triggers model retraining. Note that, unlike models trained in simulation (\Subsec{alternative_approaches}), Mowgli's model is amenable to further fine-tuning, as it is initially learned on the same distribution of environments and dynamics as the deployment setting. As a result, only minimal corrections are needed during online deployment (\eg to account for environmental drift).

Deploying \name{}'s rate control algorithm requires modifying the application's rate control logic and sending over the weights parameterizing the learned model (\Fig{policy-deployment}). Inside the application, we spawn an additional Python process responsible for serving the model. The application code and Python process communicate via an interprocess pipe; the Python process consumes live telemetry data logged by the application instrumentation code and outputs an updated bitrate. 

\subsection{Implementation}
\label{ss:implementation}
Inspired by OpenNetLab~\cite{opennetlab}, we implement \name{} on top of WebRTC~\cite{webrtc}, an open-source framework for real-time video conferencing. Our current implementation, in line with prior work~\cite{salsify}, targets unidirectional video without audio. Further, to isolate the effects of rate control in WebRTC, we set \texttt{DegradationPreference=DISABLED}~\cite{degradation_pref}.
We leverage the PyTorch~\cite{pytorch} and d3rlpy~\cite{d3rlpy} libraries to train and deploy \name{}'s learned rate control algorithm. We set the conservative loss penalty hyperparameter ($\alpha$) to 0.01 and the number of quantiles ($N$) in our distributional value representation to 128. The actor and critic neural networks have 2 hidden layers of size 256; the GRU has a hidden unit size of 32.
\vspace{-3pt}
\section{Evaluation}
\label{s:eval}

We evaluated \name{} on a diverse set of networks, both in emulation and in the wild. Our key findings are: 
\squishlist    
    \item \name{} delivers substantial QoE improvements over GCC in emulated networks, increasing average video bitrates by 15--39\% while reducing video freeze rates by 60--100\%.
    \item \name{} achieves similar performance to existing (impractical) data-driven online RL approaches, with average bitrates within 0.5--13.1\% and freeze rates within 0--19\%.
    \item \name{} can achieve wins across diverse network scenarios if sufficiently represented in consumed telemetry logs.
    \item We deploy and evaluate \name{} on real cellular networks across four U.S. cities. On target networks with high dynamism, \name{} increases average video bitrates by 17.7\% while maintaining similar levels of video freezes.
\vspace{-2pt}
    
\squishend

\begin{figure*}[t]
\centering
    \begin{subfigure}[b]{0.246\textwidth}
        \includegraphics[width=\textwidth]{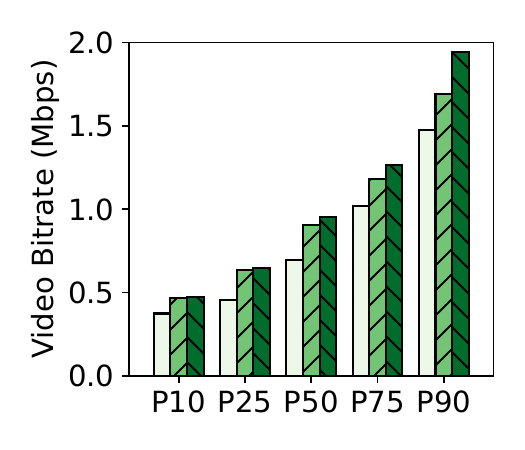}
        \vspace{-25pt}
        \caption{Video Bitrate}
    \end{subfigure}
    \begin{subfigure}[b]{0.246\textwidth}
        \includegraphics[width=\textwidth]{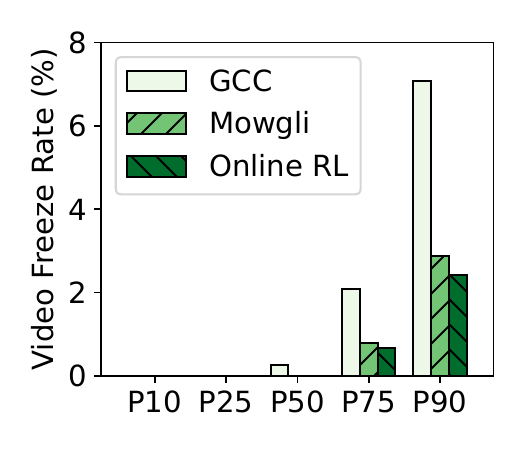}
        \vspace{-25pt}
        \caption{Video Freeze Rate}
    \end{subfigure}
    \begin{subfigure}[b]{0.246\textwidth}
        \includegraphics[width=\textwidth]{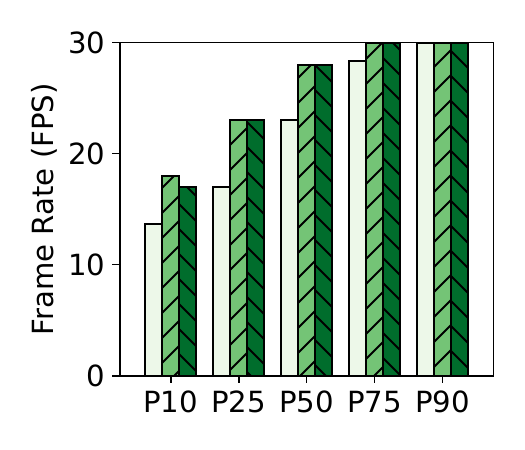}
        \vspace{-25pt}
        \caption{Frame Rate}
    \end{subfigure}
    \begin{subfigure}[b]{0.246\textwidth}
        \includegraphics[width=\textwidth]{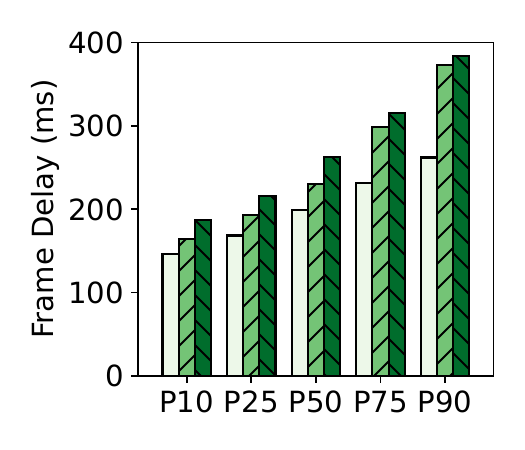}
        \vspace{-25pt}
        \caption{E2E Frame Delay}
    \end{subfigure}
    \vspace{-18pt}
    \caption{On emulated network traces, \name{} consistently outperforms GCC and nearly matches the online RL baseline while avoiding QoE degradations during training.}
    \label{fig:main_eval}
    \vspace{-15pt}
\end{figure*}

\subsection{Methodology}

\para{Experiment setup and testbed.} For lack of access to a production video conferencing deployment, we set up our own testbed to collect GCC logs and evaluate \name{}. We extended the AlphaRTC~\cite{opennetlab} fork of WebRTC to run an end-to-end client-to-client video conferencing workflow. We run both clients on a single machine and use a network emulation tool (Mahimahi~\cite{mahimahi}) to emulate the network between clients.

To evaluate \name{} on a diverse and challenging set of networks, we created a corpus of 87 hours of network bandwidth traces from two real-world datasets: FCC~\cite{fcc} broadband traces and Norway~\cite{norway} cellular traces.
We split each trace into 1-minute chunks. Following prior work~\cite{pensieve}, we filtered out traces with an average bandwidth of $<$0.2 Mbps or $>$6 Mbps. We used 60\% of the traces for training, 20\% for validation, and 20\% for testing.  Each network trace is randomly assigned to one of the following RTTs: 40, 100, and 160 ms. We use a queue length of 50 packets. 

Similar to prior work~\cite{tambur}, instead of recording and sending live video, we modify the WebRTC codebase to read from a prerecorded video. We use 9 different one-minute videos from a video conferencing dataset~\cite{video_dataset}; we randomly assign a video to each bandwidth trace. To create a corpus of ``production logs,'' we collect logs from running GCC on the network traces in the training dataset. We note that although production logs from existing video conferencing platforms~\cite{teams_dataset} exist, we do not have access to their systems to perform an evaluation; therefore, we opt to use data from our own setup.

\para{QoE metrics.} We evaluate QoE across the following metrics: (1) average received \textit{video bitrate} (Mbps), (2) \textit{video freeze rate}---fraction of session experiencing freezes (as defined by WebRTC~\cite{webrtc_eval_stats}), (3) \textit{frame rate} (FPS), and (4) average end-to-end \textit{frame delay}.
The first three metrics are already available in the WebRTC application logs.
To calculate the end-to-end frame delay, we embed a QR code into each video frame to indicate the frame ID.
We then calculate the timestamp difference between when the frame is read and when the corresponding frame is displayed. Note that we only measure the end-to-end frame delay for experiments done on emulated networks because on real networks, the clients are located on different devices and require nontrivial time synchronization.

\para{Baseline algorithms.} We compare \name{} against the following baselines:
\squishlist
    \item \textbf{Google Congestion Control (GCC)}~\cite{gcc}: the de facto rate control algorithm for video conferencing. It employs hand-tuned rule-based heuristics to characterize network behavior based on recent packet delay and loss measurements. We use the built-in implementation of GCC in WebRTC.
    \item \textbf{Online RL}: an online reinforcement learning-based approach. We implement an in-house solution following the design and methodology of prior work~\cite{loki, onrl, r3net}. The details of our implementation are provided in~\Subsec{online_rl_details}. Note that the reported results exclude quality degradations incurred during training (see \Subsec{motivations}); instead, we only present results from the model that performs the best on the test dataset.
    \item \textbf{Behavior Cloning (BC)}~\cite{widrow1964pattern, pomerleau1988alvinn}: an offline learning strategy that trains a rate control policy by imitating the behavior featured in existing logs through supervised learning.
    \item \textbf{Critic Regularized Regression (CRR)}~\cite{crr}: another offline learning strategy that relies exclusively on existing logs for training. It is the underlying mechanism for Sage~\cite{sage_cc}, a related work aimed at learning improved TCP congestion control algorithms from logs generated by dozens of different existing CC algorithms. Whereas CQL focuses on conservatively adjusting the critic function to avoid overestimating out-of-distribution actions, CRR regularizes the policy by using the critic's estimated values to guide the actor toward high-value actions of the dataset. 
  
\squishend

\subsection{Overall Performance}
\para{Main results.}
\Fig{main_eval} compares \name{} with GCC and Online RL. There are two main takeaways. First, \name{} consistently improves upon GCC; across reported percentiles, \name{} increases the average bitrate by 14.5--39.2\%, decreases the freeze rate by 59.5--100\%, and increases the frame rate by 0--35.3\%. End-to-end frame delays are within the 400 ms interactivity threshold~\cite{itu_recommendation}, even with RTTs up to 160 ms.
Second, despite learning completely offline, \name{}'s performance nearly matches that of the online RL baseline. Across percentiles, \name{} achieves an average bitrate within 0.5--13.1\% of Online RL's average. \name{}'s P75 and P90 video freeze rates are 0.77\% and 2.87\%, respectively, only slightly higher than Online RL's rates of 0.66\% and 2.41\%. For comparison, GCC's P75 and P90 freeze rates are 2.09\% and 7.09\%.

\begin{figure}[t!]
    \centering
    \begin{subfigure}[b]{0.49\columnwidth}
    \includegraphics[width=\columnwidth]{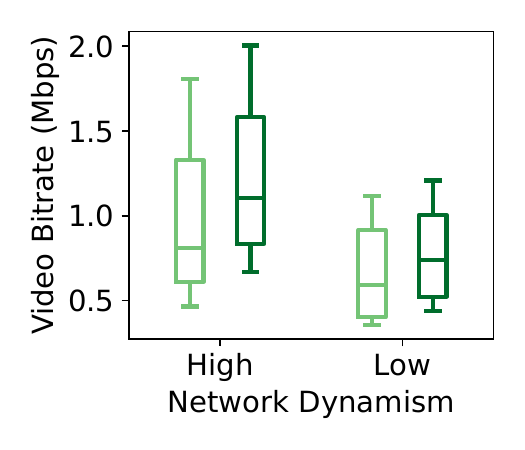}
    \vspace{-20pt}
    \caption{Video Bitrate}
    \label{fig:eval_breakdown_dynamism_0}
    \end{subfigure}
    \begin{subfigure}[b]{0.49\columnwidth}
    \includegraphics[width=\columnwidth]{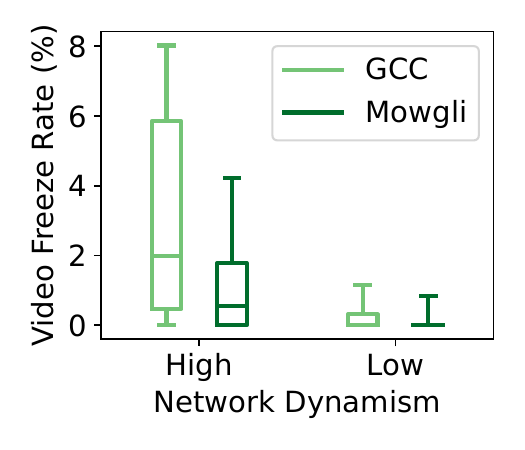}
    \vspace{-20pt}
    \caption{Freeze Rate}
    \label{fig:eval_breakdown_dynamism_1}
    \end{subfigure}
    \vspace{-5pt}
    \caption{Evaluating \name{}'s performance in network traces of varying network dynamism, measured by the degree of bandwidth variation. Whiskers show P10--P90. \name{}'s win relative to GCC is higher under high network dynamism.}
    \label{fig:eval_breakdown_dynamism}
    \vspace{-12pt}
\end{figure}

\para{Breakdown by dynamism.} To better understand how \name{} performs in the network conditions where GCC performs poorly, we split the dataset of network traces based on the degree of network dynamism. Specifically, we calculate the standard deviation of 1-second network bandwidth chunks within each trace and split the dataset along the mean standard deviation across traces. We observe that \name{} achieves its largest wins over GCC in traces with high bandwidth dynamism -- across reported percentiles, \name{} increases the average video bitrate by 10.8--43.8\% and decreases the video freeze rate by 47.4--100\% (\Fig{eval_breakdown_dynamism}). In contrast, for traces with less dynamism, \name{} increases the average video bitrate by 8.0--29.6\% and decreases the video freeze rate by 26.2--100\%.

\begin{figure}[t]
    \centering
    
    \begin{subfigure}[b]{0.49\columnwidth}
    \includegraphics[width=\columnwidth]{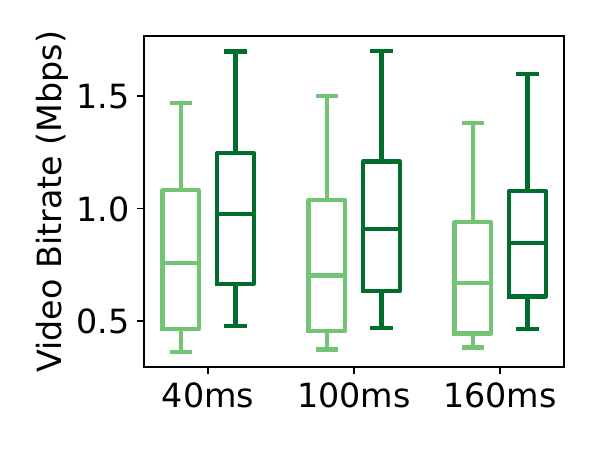}
    \vspace{-22pt}
    \caption{Video Bitrate (Diff. RTT)}
    \label{fig:eval_breakdown_0}
    \end{subfigure}
    \begin{subfigure}[b]{0.49\columnwidth}
    \includegraphics[width=\columnwidth]{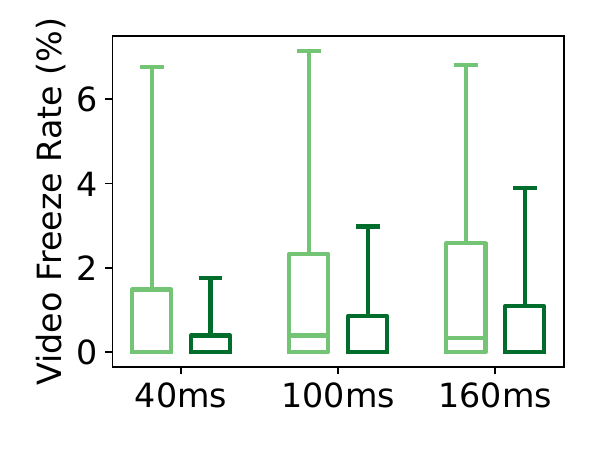}
    \vspace{-22pt}
    \caption{Freeze Rate (Diff. RTT)}
    \label{fig:fig:eval_breakdown_1}
    \end{subfigure}
    
    \begin{subfigure}[b]{0.49\columnwidth}
    \includegraphics[width=\columnwidth]{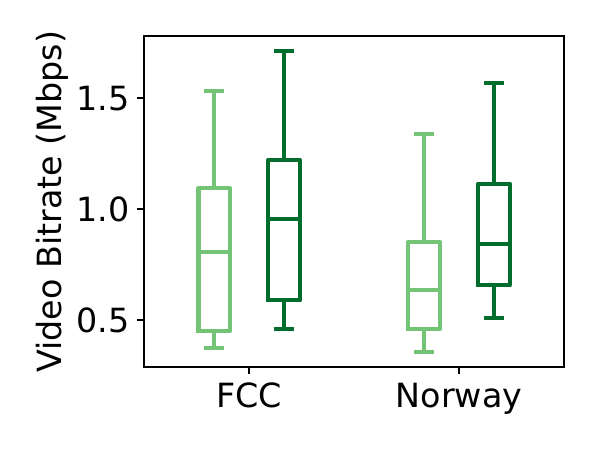}
    \vspace{-22pt}
    \caption{Video Bitrate (Diff. Dataset)}
    \label{fig:eval_breakdown_2}
    \end{subfigure}
    \begin{subfigure}[b]{0.49\columnwidth}
    \includegraphics[width=\columnwidth]{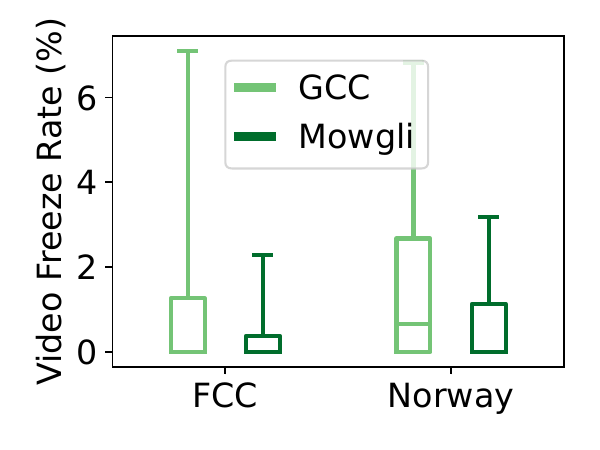}
    \vspace{-22pt}
    \caption{Freeze Rate (Diff. Dataset)}
    \label{fig:fig:eval_breakdown_3}
    \end{subfigure}
    \vspace{-5pt}
    \caption{Evaluating \name{}'s performance in network traces of varying delay (RTT) and dataset. Whiskers show P10--P90.} 
    \label{fig:eval_breakdown}
    % \vspace{-12pt}
\end{figure}

\para{Breakdown by network dataset \& delay.} In \Fig{eval_breakdown}, we break down the results from \Fig{main_eval} based on different characterizations of the underlying network traces. We find that as network delay increases, \name{}'s P50 video bitrate decreases (976 kbps $\rightarrow$ 911 kbps $\rightarrow$ 845 kbps) and its P75 video freeze rates increase (0.39\% $\rightarrow$ 0.86\% $\rightarrow$ 1.09\%). We report P75 video freeze rates because P50 video freeze rates are 0\%. 
This is expected as higher network delays increase the time it takes for a model to receive feedback about potential network congestion and react appropriately; as a result, the model is less aggressive in its bandwidth ramp-ups. When split by dataset, \name{} performs better in the FCC dataset than in the Norway dataset (954 kbps vs. 844 kbps P50 video bitrates, 0.39\% vs. 1.13\% P75 video freeze rates). The Norway dataset was collected over 3G cellular networks and therefore, features more network dynamism than the traces in the FCC dataset (which were captured on wired broadband networks).

\begin{figure}[t]
    \centering
    \includegraphics[width=.95\columnwidth]{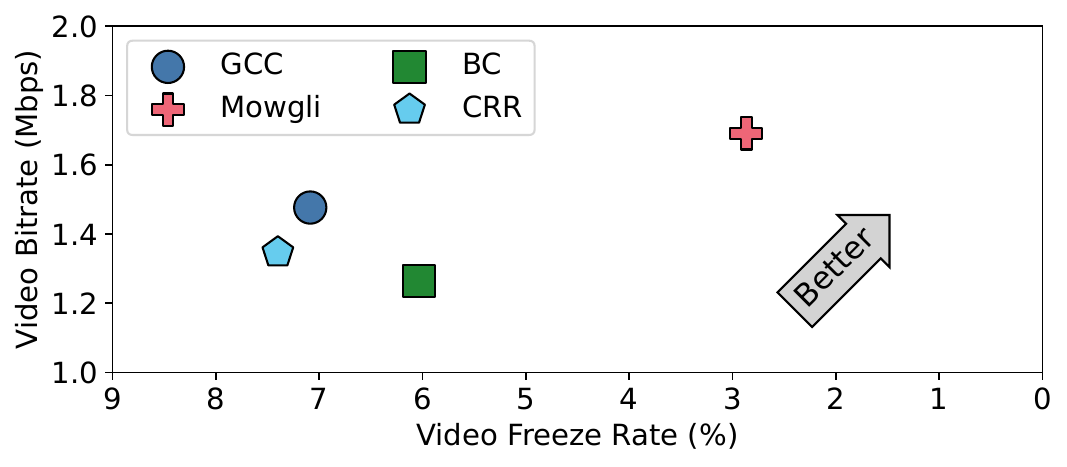}
    \vspace{-10pt}
    \caption{\name{} outperforms additional baselines in bitrate and freezes (P90 performance is shown).}
    \label{fig:other_baselines}
    \vspace{-14pt}
\end{figure}

\begin{figure}[t]
    \centering
     \begin{subfigure}[b]{0.48\columnwidth}
    \includegraphics[width=\columnwidth]{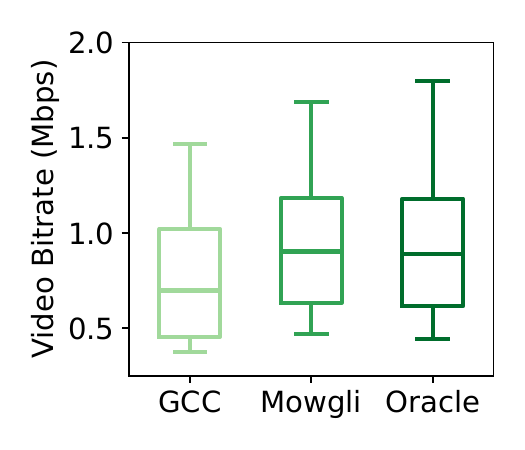}
    \vspace{-20pt}
    \caption{Video Bitrate}
    \label{fig:main_oracle_0}
    \end{subfigure}
    \begin{subfigure}[b]{0.48\columnwidth}
    \includegraphics[width=\columnwidth]{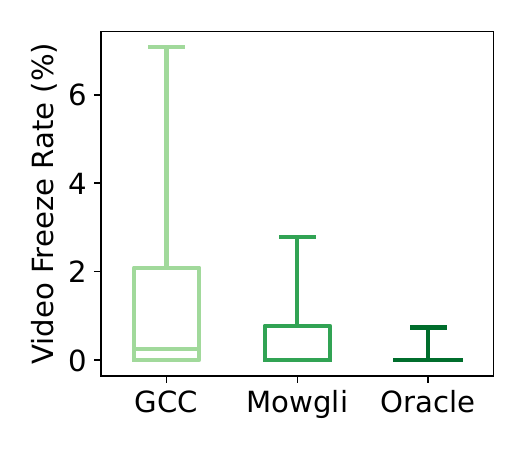}
    \vspace{-20pt}
    \caption{Freeze Rate}
    \label{fig:main_oracle_1}
    \end{subfigure}
    \vspace{-5pt}
    \caption{Comparing \name{} against an approximate oracle algorithm that represents an upper bound on performance improvements over GCC; whiskers correspond to P10 and P90.}
    \label{fig:eval_oracle}
\end{figure}

\para{Additional baselines.} In \Fig{other_baselines}, we compare \name{} with two alternative learning strategies: Behavior Cloning (BC) and Critic Regularized Regression (CRR). We find that BC behaves less aggressively than \name{}, achieving a P90 video bitrate that is 14.4\% lower than GCC, whereas \name{} increases the bitrate by 14.5\% compared with GCC.
This difference arises because BC only aims to imitate the behavior observed in the training logs and fails to effectively extrapolate to unseen scenarios. CRR, the underlying learning algorithm in Sage~\cite{sage_cc}, performs worse than GCC on both metrics, with a 4.4\% increase in the P90 video freeze rate and an 8.8\% decrease in the P90 video bitrate. We hypothesize that this is due to the lack of state-action coverage featured in the logs of Google Congestion Control; the logs of Sage, on the other hand, contain the experience of dozens of different CC algorithms reducing the likelihood of erroneous estimates.

\para{Comparison to approximate oracle.} In \Subsec{opportunity}, we implemented an approximate oracle algorithm to estimate the maximum possible improvement over GCC. This algorithm serves as an estimate of the (unattainable) upper bound on performance improvements.
\Fig{eval_oracle} compares this oracle with \name{}. Across reported percentiles, \name{} comes within 6\% of the oracle's achieved video bitrate.
Compared with GCC's video freeze rates of 2.1\% at P75 and 7.1\% at P90,
\name{} significantly reduces these rates to 0.8\% and 2.9\%,
while the oracle further lowers them to 0\% and 0.7\%.
These additional reductions are expected since the oracle algorithm has access to ground-truth network bandwidths in advance.

\begin{figure}[t]
    \centering
    \begin{subfigure}[b]{0.495\columnwidth}
    \includegraphics[width=\columnwidth]{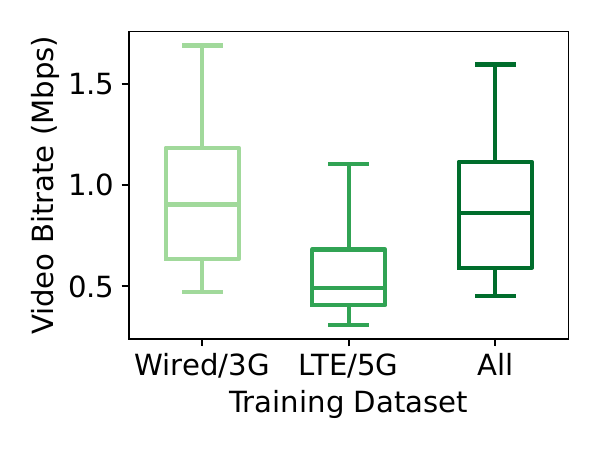}
    \vspace{-22pt}
    \caption{Video Bitrate}
    \label{fig:generalization_0}
    \end{subfigure}
    \begin{subfigure}[b]{0.495\columnwidth}
    \includegraphics[width=\columnwidth]{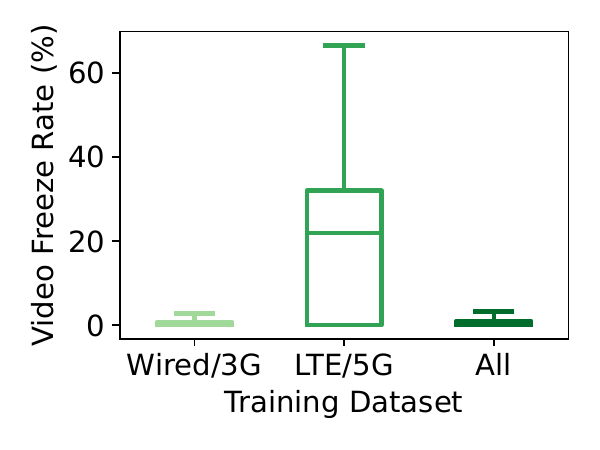}
    \vspace{-22pt}
    \caption{Freeze Rate}
    \label{fig:generalization_1}
    \end{subfigure}
    \vspace{-16pt}
    \caption{Evaluating \name{}'s performance on the Wired/3G network dataset when varying the network telemetry dataset consumed; whiskers show P10--P90.}
    \label{fig:generalization}
    \vspace{-10pt}
\end{figure}

\begin{figure}[t]
    \centering
    \begin{subfigure}[b]{0.495\columnwidth}
    \includegraphics[width=\columnwidth]{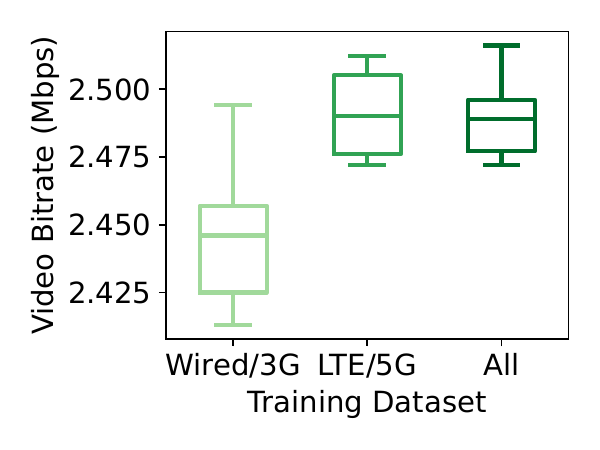}
    \vspace{-22pt}
    \caption{Video Bitrate}
    \label{fig:generalization_lte_5g_0}
    \end{subfigure}
    \begin{subfigure}[b]{0.495\columnwidth}
    \includegraphics[width=\columnwidth]{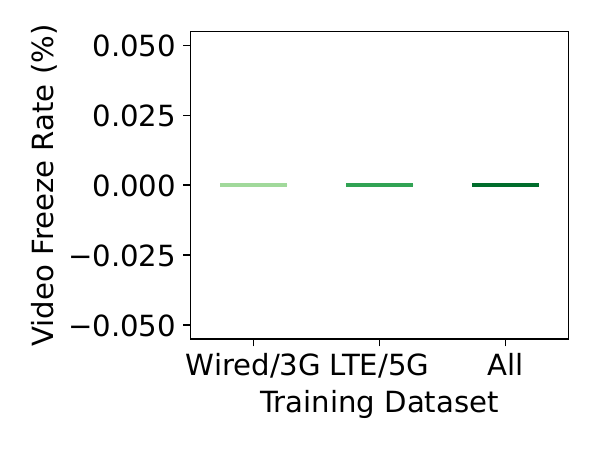}
    \vspace{-22pt}
    \caption{Freeze Rate}
    \label{fig:generalization_lte_5g_1}
    \end{subfigure}
    \vspace{-16pt}
    \caption{Evaluating \name{}'s performance on the LTE/5G network dataset when varying the content of the network telemetry dataset consumed; whiskers show P10--P90.}
    \label{fig:generalization_lte_5g}
\end{figure}

\subsection{Generalization \& Deployment Considerations}
\label{ss:generalization}
In this section, we examine the limits of \name{}’s ability to generalize to network conditions and types not represented in the production telemetry dataset. Additionally, we seek to quantify the benefits of model specialization for specific network conditions. To this end, we incorporate an additional set of network traces captured on LTE and 5G networks~\cite{ghoshal2022depth}. Following our established methodology, we collected GCC logs on these LTE/5G traces and trained a corresponding policy. We then evaluated this policy on network traces from the primary dataset, which consists of wired and 3G traces. The results are summarized in \Fig{generalization}, with two key observations.

First, we find that \name{}’s generated policies perform poorly when evaluated on network conditions that significantly differ from those featured in the training telemetry logs. For instance, the LTE/5G-trained policy performs notably worse than the Wired/3G-trained policy when evaluated on Wired/3G network traces. This results in a 45.8\% reduction in the P50 video bitrate and a 40.3$\times$ increase in the P75 video freeze rate. Importantly, this is expected due to underlying state/action distribution differences in the corresponding GCC logs (\eg GCC's average video bitrate is 1.6 Mbps higher in the LTE/5G dataset). We further discuss this in \Sec{discussion}.

Second, we observe that specializing a model for a specific set of network traces is not strictly necessary. Instead, general models can achieve strong performance across network conditions, provided they are well-represented in the training data. For example, a model trained on both network datasets performs comparably to one trained solely on the Wired/3G dataset. While the Wired/3G-specific model achieves slightly better results (a 4.6\% higher P50 video bitrate and a reduction in the P75 video freeze rate from 1.00\% to 0.77\%), the general model still performs effectively across conditions.

We conducted a similar analysis using the LTE/5G dataset, evaluating the same generalized model on these network traces (\Fig{generalization_lte_5g}). The observed trends persist, with the Wired/3G-trained policy, for example, reducing the median video bitrate by 1.8\%.

\begin{table}[t]
\centering
\small
\begin{tabular}{lll}
\toprule
\multicolumn{1}{c}{\textbf{Scenario}} & \textbf{Network} & \textbf{Cities}     \\
\midrule
\textbf{A}                              & 4G/LTE           & Princeton, NJ; San Jose, CA      \\
\textbf{B (new cities)}                 & 4G/LTE           & New York City, NY; Nashville, TN      \\
\bottomrule
\end{tabular}
\vspace{-4pt}
\caption{Cities and network types for in-the-wild evaluation.}
\label{tab:real_networks}
\vspace{-14pt}
\end{table}

\begin{figure}[t]
    \centering
    \begin{subfigure}[b]{0.47\columnwidth}
    \includegraphics[width=\columnwidth]{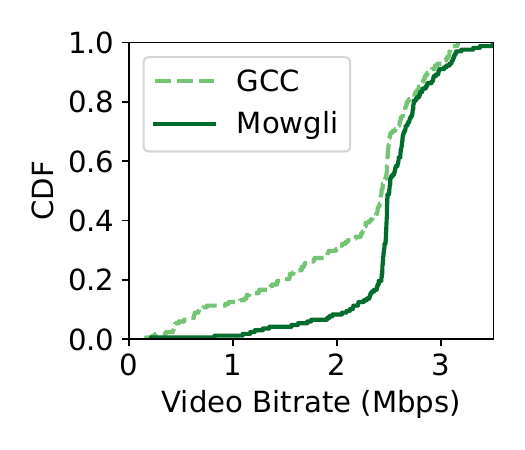}
    \vspace{-22pt}
    \caption{Scenario A}
    \label{fig:real_scenario_a}
    \end{subfigure}
    \begin{subfigure}[b]{0.47\columnwidth}
    \includegraphics[width=\columnwidth]{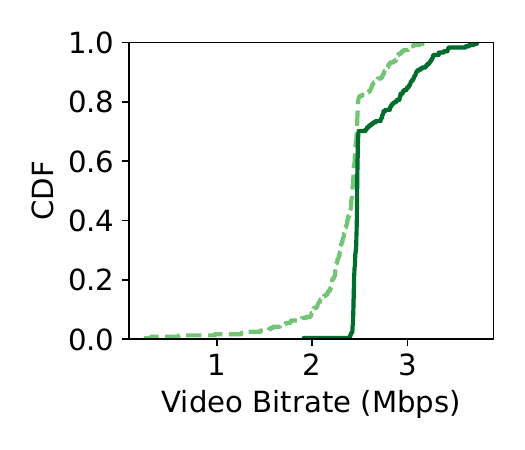}
    \vspace{-22pt}
    \caption{Scenario B}
    \label{fig:real_scenario_b}
    \end{subfigure}
    \vspace{-4pt}
    \caption{Evaluating \name{} on real-world cellular networks; scenarios are described in Table~\ref{tab:real_networks}. \name{} outperforms GCC across bitrate percentiles, while freezes remain statistically indistinguishable between the two (not shown).}
    \label{fig:real_networks}
\end{figure}

\begin{figure*}[t]
    \vspace{-20pt}
    \centering
    \begin{subfigure}[b]{0.3\textwidth}
    \includegraphics[width=\columnwidth]{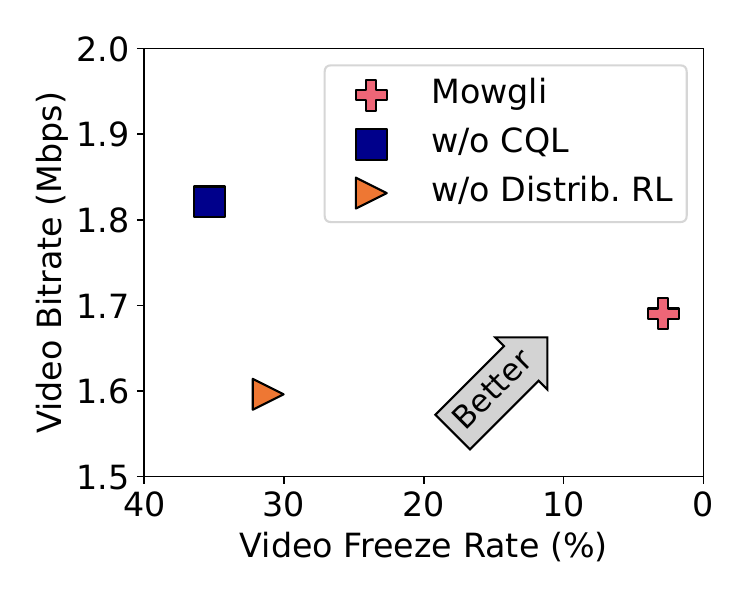}
    \vspace{-23pt}
    \caption{Algorithm Design}
    \label{fig:ablation_0}
    \end{subfigure}
    \begin{subfigure}[b]{0.3\textwidth}
    \includegraphics[width=\columnwidth]{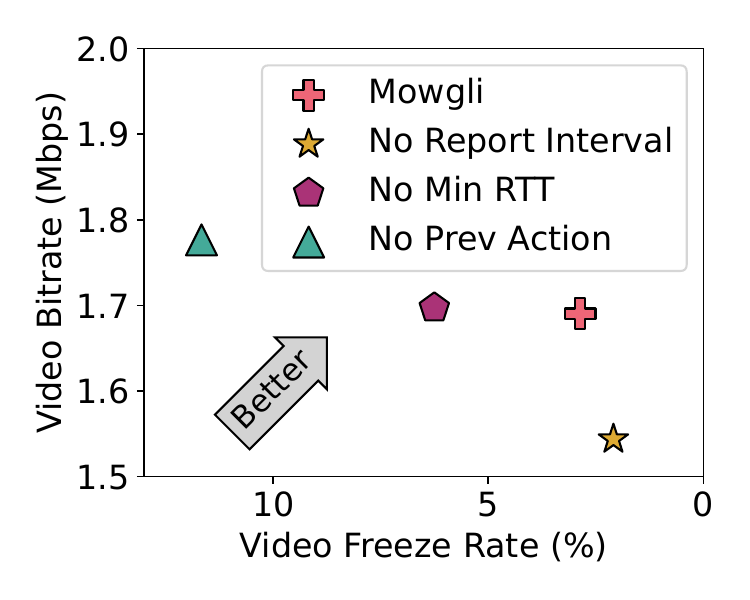}
    \vspace{-23pt}
    \caption{State Design}
    \label{fig:ablation_1}
    \end{subfigure}
    \begin{subfigure}[b]{0.3\textwidth}
    \includegraphics[width=\columnwidth]{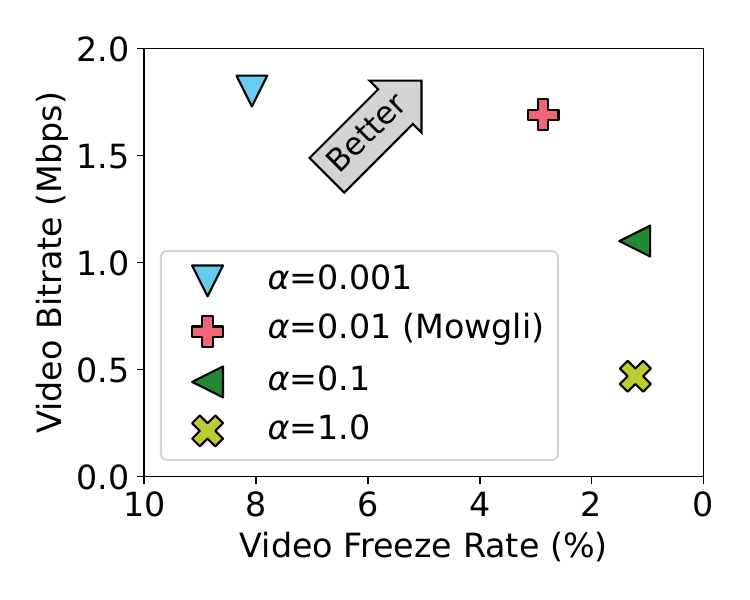}
    \vspace{-23pt}
    \caption{Alpha Value}
    \label{fig:ablation_2}
    \end{subfigure}
    \vspace{-6pt}
    \caption{Ablation results varying the algorithm design, state design, and alpha. Markers correspond to P90. \name{}'s design achieves the best bitrate-freeze tradeoff.}
    \label{fig:ablation}
    \vspace{-18pt}
\end{figure*}

\subsection{Real World Experiments}
We evaluated \name{} and GCC on real cellular networks in 4 different cities (\Tab{real_networks}). In these experiments, we set up conferencing sessions with a client running on a Macbook Pro (tethered to a Google Pixel) and a server in the cloud. Experiments were performed in several mobility scenarios (\eg train, bus, car, walking, and stationary). To generate a dataset of GCC logs, we collected over 8 hours of video calls using GCC in Princeton, NJ and San Jose, CA on a 4G/LTE network. Once \name{} generated a policy, we evaluated the resulting model in 2 different scenarios: (a) the same networks and cities and (b) different cities. During our evaluation, we alternated running GCC and the generated policy, collecting over 4 hours of data for each scenario.

We observe that \name{}'s wins over GCC extend to cellular networks in both scenarios, increasing video bitrates across reported percentiles by 3.0\%--2.1$\times$ (\Fig{real_scenario_a}) and 2.0\%--20.8\% \Fig{real_scenario_b}).
Video freezes---as a rare event---are inherently challenging to measure reliably~\cite{puffer}.
While we currently lack sufficient evidence to draw definitive conclusions,
the observed rates across 120 runs per policy/scenario appear statistically indistinguishable 
(results not shown). 

\subsection{Ablation Studies and Microbenchmarks}
\label{ss:ablation}

\para{Algorithm design ablation.} \Fig{ablation_0} compares \name{} with two variants: (1) \name{} without the conservative learning regularizer, and (2) \name{} without the distributional representation. Removing the regularizer, which adjusts the learned estimates of the critic function based on sample confidence, makes the model prone to actions with erroneously high estimated values and results in 11.3$\times$ higher P90 video freeze rate. Without the distributional representation, which accounts for external phenomena outside the control of the bitrate decision-making policy, P90 video bitrates drop by 5.6\% and P90 video freeze rates increase by 9.9$\times$.

\para{Varying state design.} \Fig{ablation_1} shows the benefits of the additional state features. Removing ``Report Intervals'', which indicates the staleness of the sender-side transport feedback report, leads to an 8.7\% lower video bitrate. Removing ``Min RTT'', which indicates how fast the client can receive (and react to) feedback and helps control the model's aggression, leads to a 1.2$\times$ higher freeze rate. Removing ``Prev Action'', which enables smooth rate control, results in a 3.1$\times$ increase in the video freeze rate.

\para{CQL $\alpha$ parameter sensitivity.} $\alpha$ dictates the relative weight of the conservative penalty in the learning algorithm. In \Fig{ablation_2}, a larger $\alpha$ %(e.g. 0.1, 1.0) 
creates a conservative, low-risk policy ($\sim$57\% lower freeze rates), but lowers video bitrates by 34.9\% and 72.3\%, respectively. In contrast, $\alpha < 0.01$ increases the amount of deviation (and therefore risks: 1.8$\times$ higher video freeze rate) albeit increasing bitrate by 6.6\%.

\para{System overheads.} We study the compute and storage overhead of deploying \name{}. The compressed (state, action, reward) logs take $\sim$117 kB for a 1-minute call. \name{}'s generated policy (weights) is 316 kB (corresponding to 79k parameters) and takes $\sim$6 ms to run on the CPU. 
\vspace{-2pt}

\section{Related Work}
\label{s:related}

\para{Alternative designs and knobs for video conferencing.} Recent work has explored different optimization dimensions in videoconferencing. Salsify~\cite{salsify} co-designs the video codec and transport protocol to quickly respond to changing network conditions and achieve low latency. Gemino~\cite{gemino} designs a neural codec for high perceptual quality in low bandwidth environments. Grace~\cite{grace} designs a loss-resilient neural codec. AFR~\cite{afr} adapts the frame rate for video conferencing applications with ultra-high-definition demand. \name{} is complementary to these works and can be extended to incorporate these alternate dimensions, which we leave for future work. 

\para{ML for networked systems.} Machine learning has been applied to various other challenges in networked systems. For example, to improve adaptive bitrate (ABR) selection in video streaming, Pensieve~\cite{pensieve} trains a reinforcement learning agent in simulation, whereas Puffer~\cite{puffer} deploys a neural network-based model predictive control scheme under real-world conditions. Unlike video streaming, which deals with discrete bitrate adaptation, video conferencing requires continuous rate control, tighter latency constraints, and frequent (every 50 ms) decision-making for on-the-fly encoding and compression. Beyond ABR, ML has been explored for TCP congestion control; Orca~\cite{orca} employs an online RL agent, while Sage~\cite{sage_cc} trains offline but requires additional data collection with multiple expert policies. In contrast, \name{} leverages existing logs of a single policy (\eg GCC).

\section{Discussion}
\label{s:discussion}

\para{Pitfalls of leveraging production traces.} Conceptually, the challenges described in \Subsec{design_challenges} align with those in recent research on counterfactual reasoning in trace-driven simulations for adaptive bitrate selection~\cite{veritas, causalsim, data_driven_networking_biases, video_streaming_confounders}, \eg the impact of latent confounders (factors that impact system dynamics but are not explicitly featured in logs). Veritas~\cite{veritas} addresses this by employing advanced statistical methods to explicitly infer these variables. While these efforts aim to refine simulations for more reliable counterfactual reasoning, \name{} focuses on generating an enhanced rate control policy derived from the behavior of an existing heuristic algorithm. Rather than depending on accurate counterfactual reasoning universally, \name{} limits it to scenarios where sufficient data is available for reliable analysis (\ie a conservative approach---\Subsec{generator}). In cases lacking enough log data, \name{} learns to mimic the baseline GCC behavior. Incorporating insights from these works into \name{} and other video conferencing systems is a promising avenue for future exploration, such as improving the critic by explicitly inferring latent confounders.

\para{Limitations of a data-driven approach.} A limitation of any data-driven approach is that if the underlying data distribution shifts substantially from the one captured in training, the model must be updated~\cite{data_drift}. In the case of \name{}, encountering previously unseen network conditions (\eg new bandwidth ranges) would necessitate a model update (\Subsec{generalization}). This limitation reflects the inherent trade-off of data-driven methods: \name{} is constrained to perform well in environments it has been trained on, unlike GCC, which operates under a broader set of pre-designed assumptions. However, unlike models trained in simulation or from scratch, \name{}’s model---trained on real user interactions and dynamics---starts on a much closer distribution to its deployment setting. This makes it more amenable to fine-tuning, requiring only minimal adjustments to account for environmental drift.
\section{Conclusion}
\label{s:conclusion}
In this paper, we presented \name{}, a system for \emph{practically} learning improved bitrate control algorithms for video conferencing. Whereas existing data-driven reinforcement learning-based approaches fall short of meeting the practicality constraints of production settings, \name{} demonstrates the possibility of learning from data already collected in existing system telemetry logs (and avoiding the QoE disruptions of prior work) to generate improved bitrate control algorithms.

\para{Acknowledgements.} We thank our shepherd, Zahaib Akhtar, and the anonymous NSDI reviewers for their valuable feedback and comments. This work was supported by NSF CNS grants 2147909, 2151630, 2140552, 2153449, and 2152313.
\label{last}
\balance
\Urlmuskip=0mu plus 1mu\relax
\bibliographystyle{abbrv}
\bibliography{paper}

\begin{thebibliography}{10}

\bibitem{zoom_logging2}
{Accessing Zoom Meeting and Phone Statistics}.
\newblock \url{https://support.zoom.com/hc/en/article?id=zm_kb&sysparm_article=KB0070504}.

\bibitem{teams_dataset}
{Bandwidth Estimation Challenge Dataset}.
\newblock \url{https://www.microsoft.com/en-us/research/academic-program/bandwidth-estimation-challenge/data}.

\bibitem{fcc}
{Measuring Broadband America}.
\newblock \url{https://www.fcc.gov/general/measuring-broadband-america}.

\bibitem{teams_logging2}
{Microsoft Teams Client Diagnostic Logs}.
\newblock \url{https://learn.microsoft.com/en-us/microsoftteams/log-files}.

\bibitem{degradation_pref}
{RTP Parameters}.
\newblock \url{https://source.chromium.org/chromium/chromium/src/+/main:third_party/webrtc/api/rtp_parameters.h}.

\bibitem{zoom_production_logging}
{Understanding the Consent Settings in Data \& Privacy Center}.
\newblock \url{https://support.zoom.com/hc/en/article?id=zm_kb&sysparm_article=KB0057779}.

\bibitem{teams_logging}
{Use real-time telemetry to troubleshoot poor meeting quality}.
\newblock \url{https://learn.microsoft.com/en-us/microsoftteams/use-real-time-telemetry-to-troubleshoot-poor-meeting-quality}.

\bibitem{webrtc}
{WebRTC}.
\newblock \url{https://webrtc.org/}.

\bibitem{webrtc_logging}
{WebRTC Logging}.
\newblock \url{https://webrtc.github.io/webrtc-org/native-code/logging/}.

\bibitem{zoom_logging}
{Zoom Service Quality Logging}.
\newblock \url{https://developers.zoom.us/docs/video-sdk/web/quality/}.

\bibitem{itu_recommendation}
{G.114 One-way transmission time}.
\newblock Technical report, International Telecommunication Union, May 2003.

\bibitem{sim2realitygap2}
{Closing the Simulation-to-Reality Gap for Deep Robotic Learning}, 2017.
\newblock \url{https://research.google/blog/closing-the-simulation-to-reality-gap-for-deep-robotic-learning/}.

\bibitem{webrtc_eval_stats}
{Identifiers for WebRTC's Statistics API}, July 2018.
\newblock \url{https://www.w3.org/TR/webrtc-stats/}.

\bibitem{orca}
S.~Abbasloo, C.-Y. Yen, and H.~J. Chao.
\newblock {Classic Meets Modern: a Pragmatic Learning-Based Congestion Control for the Internet}.
\newblock In {\em Proceedings of the Annual conference of the ACM Special Interest Group on Data Communication on the applications, technologies, architectures, and protocols for computer communication}, pages 632--647, 2020.

\bibitem{oboe}
Z.~Akhtar, Y.~S. Nam, R.~Govindan, S.~Rao, J.~Chen, E.~Katz-Bassett, B.~Ribeiro, J.~Zhan, and H.~Zhang.
\newblock Oboe: Auto-tuning video abr algorithms to network conditions.
\newblock In {\em Proceedings of the 2018 Conference of the ACM Special Interest Group on Data Communication}, pages 44--58, 2018.

\bibitem{causalsim}
A.~Alomar, P.~Hamadanian, A.~Nasr-Esfahany, A.~Agarwal, M.~Alizadeh, and D.~Shah.
\newblock {CausalSim: A Causal Framework for Unbiased Trace-Driven Simulation}.
\newblock In {\em 20th USENIX Symposium on Networked Systems Design and Implementation (NSDI 23)}, pages 1115--1147, 2023.

\bibitem{data_driven_networking_biases}
M.~Bartulovic, J.~Jiang, S.~Balakrishnan, V.~Sekar, and B.~Sinopoli.
\newblock {Biases in Data-Driven Networking, and What to Do About Them}.
\newblock In {\em Proceedings of the 16th ACM Workshop on Hot Topics in Networks}, pages 192--198, 2017.

\bibitem{distributional_rl}
M.~G. Bellemare, W.~Dabney, and M.~Rowland.
\newblock {\em {Distributional Reinforcement Learning}}.
\newblock MIT Press, 2023.

\bibitem{veritas}
C.~Bothra, J.~Gao, S.~Rao, and B.~Ribeiro.
\newblock {Veritas: Answering causal queries from video streaming traces}.
\newblock In {\em Proceedings of the ACM SIGCOMM 2023 Conference}, pages 738--753, 2023.

\bibitem{sim2realitygap1}
K.~Bousmalis, A.~Irpan, P.~Wohlhart, Y.~Bai, M.~Kelcey, M.~Kalakrishnan, L.~Downs, J.~Ibarz, P.~Pastor, K.~Konolige, S.~Levine, and V.~Vanhoucke.
\newblock {Using Simulation and Domain Adaptation to Improve Efficiency of Deep Robotic Grasping}.
\newblock In {\em 2018 IEEE International Conference on Robotics and Automation (ICRA)}, page 4243–4250. IEEE Press, 2018.

\bibitem{gcc}
G.~Carlucci, L.~De~Cicco, S.~Holmer, and S.~Mascolo.
\newblock {Analysis and Design of the Google Congestion Control for Web Real-time Communication (WebRTC)}.
\newblock In {\em Proceedings of the 7th International Conference on Multimedia Systems}, pages 1--12, 2016.

\bibitem{grace}
Y.~Cheng, Z.~Zhang, H.~Li, A.~Arapin, Y.~Zhang, Q.~Zhang, Y.~Liu, K.~Du, X.~Zhang, F.~Y. Yan, et~al.
\newblock Grace: Loss-resilient real-time video through neural codecs.
\newblock In {\em 21st USENIX Symposium on Networked Systems Design and Implementation (NSDI '24)}, pages 509--531, 2024.

\bibitem{gru}
K.~Cho, B.~van Merrienboer, {\c{C}}.~G{\"{u}}l{\c{c}}ehre, F.~Bougares, H.~Schwenk, and Y.~Bengio.
\newblock {Learning Phrase Representations using {RNN} Encoder-Decoder for Statistical Machine Translation}.
\newblock {\em CoRR}, abs/1406.1078, 2014.

\bibitem{quantile_regression}
W.~Dabney, M.~Rowland, M.~Bellemare, and R.~Munos.
\newblock {Distributional Reinforcement Learning with Quantile Regression}.
\newblock In {\em Proceedings of the AAAI Conference on Artificial Intelligence}, volume~32, 2018.

\bibitem{cloudlab}
D.~Duplyakin, R.~Ricci, A.~Maricq, G.~Wong, J.~Duerig, E.~Eide, L.~Stoller, M.~Hibler, D.~Johnson, K.~Webb, A.~Akella, K.~Wang, G.~Ricart, L.~Landweber, C.~Elliott, M.~Zink, E.~Cecchet, S.~Kar, and P.~Mishra.
\newblock {The Design and Operation of {CloudLab}}.
\newblock In {\em 2019 USENIX Annual Technical Conference (USENIX ATC 19)}, pages 1--14, 2019.

\bibitem{opennetlab}
J.~Eo, Z.~Niu, W.~Cheng, F.~Y. Yan, R.~Gao, J.~Kardhashi, S.~Inglis, M.~Revow, B.-G. Chun, P.~Cheng, et~al.
\newblock {OpenNetLab: Open platform for RL-based Congestion Control for Real-Time Communications}.
\newblock {\em Proc. of APNet}, 2022.

\bibitem{r3net}
J.~Fang, M.~Ellis, B.~Li, S.~Liu, Y.~Hosseinkashi, M.~Revow, A.~Sadovnikov, Z.~Liu, P.~Cheng, S.~Ashok, et~al.
\newblock {Reinforcement Learning for Bandwidth Estimation and Congestion Control in Real-Time Communications}.
\newblock {\em arXiv preprint arXiv:1912.02222}, 2019.

\bibitem{salsify}
S.~Fouladi, J.~Emmons, E.~Orbay, C.~Wu, R.~S. Wahby, and K.~Winstein.
\newblock {Salsify: Low-Latency Network Video through Tighter Integration between a Video Codec and a Transport Protocol}.
\newblock In {\em 15th USENIX Symposium on Networked Systems Design and Implementation (NSDI '18)}, pages 267--282, 2018.

\bibitem{ghoshal2022depth}
M.~Ghoshal, Z.~J. Kong, Q.~Xu, Z.~Lu, S.~Aggarwal, I.~Khan, Y.~Li, Y.~C. Hu, and D.~Koutsonikolas.
\newblock {An In-Depth Study of Uplink Performance of 5G mmWave Networks}.
\newblock In {\em Proceedings of the ACM SIGCOMM Workshop on 5G and Beyond Network Measurements, Modeling, and Use Cases}, pages 29--35, 2022.

\bibitem{sac}
T.~Haarnoja, A.~Zhou, P.~Abbeel, and S.~Levine.
\newblock {Soft Actor-Critic: Off-Policy Maximum Entropy Deep Reinforcement Learning with a Stochastic Actor}.
\newblock In {\em International conference on machine learning}, pages 1861--1870. PMLR, 2018.

\bibitem{imitation_learning}
A.~Hussein, M.~M. Gaber, E.~Elyan, and C.~Jayne.
\newblock {Imitation Learning: A Survey of Learning Methods}.
\newblock {\em ACM Computing Surveys (CSUR)}, 50(2):1--35, 2017.

\bibitem{cql}
A.~Kumar, A.~Zhou, G.~Tucker, and S.~Levine.
\newblock {Conservative Q-Learning for Offline Reinforcement Learning}.
\newblock {\em Advances in Neural Information Processing Systems}, 33:1179--1191, 2020.

\bibitem{levine_offline_rl_tutorial}
S.~Levine, A.~Kumar, G.~Tucker, and J.~Fu.
\newblock {Offline Reinforcement Learning: Tutorial, Review, and Perspectives on Open Problems}.
\newblock {\em arXiv preprint arXiv:2005.01643}, 2020.

\bibitem{data_drift}
J.~Lu, A.~Liu, F.~Dong, F.~Gu, J.~Gama, and G.~Zhang.
\newblock {Learning under Concept Drift: A Review}.
\newblock {\em IEEE Transactions on Knowledge and Data Engineering}, 2018.

\bibitem{pensieve}
H.~Mao, R.~Netravali, and M.~Alizadeh.
\newblock {Neural Adaptive Video Streaming with Pensieve}.
\newblock In {\em Proceedings of the conference of the ACM special interest group on data communication}, pages 197--210, 2017.

\bibitem{afr}
Z.~Meng, T.~Wang, Y.~Shen, B.~Wang, M.~Xu, R.~Han, H.~Liu, V.~Arun, H.~Hu, and X.~Wei.
\newblock {Enabling High Quality Real-Time Communications with Adaptive Frame-Rate}.
\newblock In {\em 20th USENIX Symposium on Networked Systems Design and Implementation (NSDI '23)}, pages 1429--1450, 2023.

\bibitem{mahimahi}
R.~Netravali, A.~Sivaraman, S.~Das, A.~Goyal, K.~Winstein, J.~Mickens, and H.~Balakrishnan.
\newblock {Mahimahi: Accurate Record-and-Replay for HTTP}.
\newblock In {\em 2015 USENIX Annual Technical Conference (USENIX ATC '15)}, pages 417--429, 2015.

\bibitem{video_dataset}
E.~Nowroozi, A.~Dehghantanha, R.~M. Parizi, and K.-K.~R. Choo.
\newblock {A Survey of Machine Learning Techniques in Adversarial Image Forensics}.
\newblock {\em Computers \& Security}, page 102092, 2020.

\bibitem{pytorch}
A.~Paszke, S.~Gross, F.~Massa, A.~Lerer, J.~Bradbury, G.~Chanan, T.~Killeen, Z.~Lin, N.~Gimelshein, L.~Antiga, et~al.
\newblock {Pytorch: An imperative style, high-performance deep learning library}.
\newblock {\em Advances in neural information processing systems}, 32, 2019.

\bibitem{pomerleau1988alvinn}
D.~A. Pomerleau.
\newblock {Alvinn: An autonomous land vehicle in a neural network}.
\newblock {\em Advances in neural information processing systems}, 1, 1988.

\bibitem{stable_baselines3}
A.~Raffin, A.~Hill, A.~Gleave, A.~Kanervisto, M.~Ernestus, and N.~Dormann.
\newblock {Stable-Baselines3: Reliable Reinforcement Learning Implementations}.
\newblock {\em The Journal of Machine Learning Research}, 22(1):12348--12355, 2021.

\bibitem{norway}
H.~Riiser, P.~Vigmostad, C.~Griwodz, and P.~Halvorsen.
\newblock {Commute path bandwidth traces from 3G networks: analysis and applications}.
\newblock In {\em Proceedings of the 4th ACM Multimedia Systems Conference}, pages 114--118, 2013.

\bibitem{tambur}
M.~Rudow, F.~Y. Yan, A.~Kumar, G.~Ananthanarayanan, M.~Ellis, and K.~Rashmi.
\newblock {Tambur: Efficient loss recovery for videoconferencing via streaming codes}.
\newblock In {\em 20th USENIX Symposium on Networked Systems Design and Implementation (NSDI '23)}, pages 953--971, 2023.

\bibitem{d3rlpy}
T.~Seno and M.~Imai.
\newblock {d3rlpy: An offline deep reinforcement learning library}.
\newblock {\em The Journal of Machine Learning Research}, 23(1):14205--14224, 2022.

\bibitem{gemino}
V.~Sivaraman, P.~Karimi, V.~Venkatapathy, M.~Khani, S.~Fouladi, M.~Alizadeh, F.~Durand, and V.~Sze.
\newblock {Gemino: Practical and Robust Neural Compression for Video Conferencing}.
\newblock In {\em 21st USENIX Symposium on Networked Systems Design and Implementation (NSDI '24)}, pages 569--590, 2024.

\bibitem{video_streaming_confounders}
P.~C. Sruthi, S.~Rao, and B.~Ribeiro.
\newblock {Pitfalls of data-driven networking: A case study of latent causal confounders in video streaming}.
\newblock In {\em Proceedings of the Workshop on Network Meets AI \& ML}, NetAI '20, page 42–47, New York, NY, USA, 2020. Association for Computing Machinery.

\bibitem{crr}
Z.~Wang, A.~Novikov, K.~Zolna, J.~S. Merel, J.~T. Springenberg, S.~E. Reed, B.~Shahriari, N.~Siegel, C.~Gulcehre, N.~Heess, and N.~de~Freitas.
\newblock {Critic Regularized Regression}.
\newblock In {\em Advances in Neural Information Processing Systems}, volume~33, pages 7768--7778, 2020.

\bibitem{watkins1992q}
C.~J. Watkins and P.~Dayan.
\newblock Q-learning.
\newblock {\em Machine learning}, 8:279--292, 1992.

\bibitem{widrow1964pattern}
B.~Widrow.
\newblock {Pattern recognition and adaptive control}.
\newblock {\em IEEE Transactions on Applications and Industry}, 83(74):269--277, 1964.

\bibitem{sprout}
K.~Winstein, A.~Sivaraman, and H.~Balakrishnan.
\newblock {Stochastic Forecasts Achieve High Throughput and Low Delay over Cellular Networks}.
\newblock In {\em 10th USENIX Symposium on Networked Systems Design and Implementation (NSDI '13)}, pages 459--471, 2013.

\bibitem{puffer}
F.~Y. Yan, H.~Ayers, C.~Zhu, S.~Fouladi, J.~Hong, K.~Zhang, P.~Levis, and K.~Winstein.
\newblock {Learning in Situ: a Randomized Experiment in Video Streaming}.
\newblock In {\em 17th USENIX Symposium on Networked Systems Design and Implementation (NSDI '20)}, pages 495--511, 2020.

\bibitem{sage_cc}
C.-Y. Yen, S.~Abbasloo, and H.~J. Chao.
\newblock {Computers Can Learn from the Heuristic Designs and Master Internet Congestion Control}.
\newblock In {\em Proceedings of the ACM SIGCOMM 2023 Conference}, pages 255--274, 2023.

\bibitem{verus}
Y.~Zaki, T.~P{\"o}tsch, J.~Chen, L.~Subramanian, and C.~G{\"o}rg.
\newblock {Adaptive congestion control for unpredictable cellular networks}.
\newblock In {\em Proceedings of the 2015 ACM Conference on Special Interest Group on Data Communication}, pages 509--522, 2015.

\bibitem{loki}
H.~Zhang, A.~Zhou, Y.~Hu, C.~Li, G.~Wang, X.~Zhang, H.~Ma, L.~Wu, A.~Chen, and C.~Wu.
\newblock {Loki: Improving Long Tail Performance of Learning-Based Real-Time Video Adaptation by Fusing Rule-Based Models}.
\newblock In {\em Proceedings of the 27th Annual International Conference on Mobile Computing and Networking}, pages 775--788, 2021.

\bibitem{onrl}
H.~Zhang, A.~Zhou, J.~Lu, R.~Ma, Y.~Hu, C.~Li, X.~Zhang, H.~Ma, and X.~Chen.
\newblock {OnRL: Improving Mobile Video Telephony Via Online Reinforcement Learning}.
\newblock In {\em Proceedings of the 26th Annual International Conference on Mobile Computing and Networking}, pages 1--14, 2020.

\bibitem{concerto}
A.~Zhou, H.~Zhang, G.~Su, L.~Wu, R.~Ma, Z.~Meng, X.~Zhang, X.~Xie, H.~Ma, and X.~Chen.
\newblock {Learning to Coordinate Video Codec with Transport Protocol for Mobile Video Telephony}.
\newblock In {\em The 25th Annual International Conference on Mobile Computing and Networking}, pages 1--16, 2019.

\end{thebibliography}

\end{sloppypar}

\appendix

\newpage
\twocolumn

\section{Appendix}
\label{s:appendix}

\subsection{Online RL Setup}
\label{ss:online_rl_details}

\para{Experiment Setup and Testbed.} Similar to prior work~\cite{loki, orca}, our online RL baseline relies on a cloud server to support the training. The centralized server has an AMD EPYC 7543P 32-core CPU, 256 GB of RAM, and an NVIDIA RTX A6000-48G GPU. We use 30 CloudLab~\cite{cloudlab} nodes (``workers'') to emulate end users in our experiments, each with an Intel Xeon D-1548 8-core CPU and 64 GB of RAM.  We modify Stable Baselines3 v2.1.0~\cite{stable_baselines3} to support ingesting state-action-reward tuples from parsed traces 
and use PyTorch v2.0.1 to train and serve the neural network. Similar to prior work, we use a state-of-the-art off-policy algorithm~\cite{sac}. Following OnRL~\cite{onrl}, we implement a fallback mechanism that allows the sender to temporarily downgrade to the default heuristic (GCC) if catastrophic QoE degradations are detected during training.

In every round of online RL training, the RL server dispatches the latest model to all 30 worker nodes. Similar to how we collect GCC logs in our main experiments, we run both the sender and receiver client on the same node. Each worker uses Mahimahi~\cite{mahimahi} to replay a random one-minute network bandwidth trace from the same diverse dataset \name{} uses. At the end of every video conferencing session, each node parses the sender/receiver logs to construct a state-action-reward trace and sends it back to the RL server. The RL server performs one training epoch based on the aggregated state-action-reward tuples across all workers and repeats this process. We enumerate the hyperparameters used by our online RL baseline in \Tab{online_rl_hyperparams}.

We retain the state and action formulation as defined in the main text. However, we observe that using the following reward definition for training the online RL policy further improves performance:

\begin{equation}
    \begin{split}
    R & = throughput \cdot delay \cdot (1 - \gamma \cdot loss) \\ 
      & - \zeta \cdot \max(prev\_action - sending\_bitrate, 0) \\
      & - use\_gcc \cdot gcc\_penalty
    \end{split}
\end{equation}

where $\gamma=2$, and $\zeta=3$. $use\_gcc$ acts as an indicator variable for whether the fallback mechanism was invoked; we set $gcc\_penalty=0.05$. We normalize $throughput$, $delay$, $loss$, $prev\_action$, and $sending\_bitrate$ to (0, 4.5 Mbps), (0, 1000 ms), (0, 1), (0, 4.5 Mbps), and (0, 4.5 Mbps).

\begin{table}[t]
    \centering
    \begin{tabular}{cc}
        \toprule
        \textbf{Hyperparameter} & \textbf{Value} \\
        \midrule
        Learning Rate & 5e-5 \\
        Batch Size & 512 \\
        Gradient Steps & 500 \\
        Replay Buffer Size & 1e6 \\
        Init. Entropy Coefficient & 0.5 \\
        GRU Hidden Size & 32 \\
        Num Parallel Workers & 30 \\
        Optimizer & Adam \\
        \bottomrule
    \end{tabular}
    \caption{Online RL Hyperparameter Values. For all other unspecified hyperparameters, we use the default values of the implementation in Stable Baselines3~\cite{stable_baselines3}. The same learning rate is used for all networks (Q-Values, Actor, and Value function).}
    \label{tab:online_rl_hyperparams}
\end{table}

\end{document}